\pdfoutput=1

\documentclass[journal]{IEEEtran}
\ifCLASSINFOpdf
\else
\fi

\usepackage{amstext,amsmath,amssymb}
\usepackage{amsthm}
\usepackage{times}
\usepackage{graphicx}
\usepackage{epstopdf}
\usepackage{algorithmic}
\usepackage{algorithm}
\usepackage{url}
\usepackage{graphicx}
\usepackage{color}
\usepackage{verbatim}
\usepackage{cite}
\usepackage{comment}
\usepackage{multicol, blindtext}
\usepackage{multirow}
\usepackage{subfig}
\usepackage{float}
\usepackage{soul}
\soulregister\cite7
\soulregister\ref7

\newcommand{\bdm}{
    \begin{displaymath}}

\newcommand{\edm}{
    \end{displaymath}}

\newcommand{\be}{
    \begin{equation}}

\newcommand{\ee}{
    \end{equation}}

\newcommand{\bea}{
    \begin{eqnarray}}

\newcommand{\eea}{
    \end{eqnarray}}

\newcommand{\beit}{\begin{itemize}}
\newcommand{\eeit}{\end{itemize}}

\newcommand{\eat}[1]{ }

\begin{document}
%
\title{Workflow Scheduling in the Cloud with Weighted Upward-rank Priority Scheme Using Random Walk
and Uniform Spare Budget Splitting}
%
%
%

\author{Hang~Zhang, Xiaoying~Zheng\IEEEauthorrefmark{1}, Ye~Xia, and~Mingqi~Li
\thanks{\IEEEauthorrefmark{1}Corresponding author}
\thanks{H. Zhang is with the School of Computer Engineering and Science, Shanghai University, Shanghai, 200444, China.H. Zhang is also with Shanghai Advanced Research Institute, Chinese Academy of Sciences, Shanghai, 201210, China. E-mail: (hang129@163.com)}
\thanks{X. Zheng and M. Li are with Shanghai Advanced Research Institute, Chinese Academy of Sciences, Shanghai, 201210, China.  E-mail: ({zhengxy, limq}@sari.ac.cn)}
\thanks{Y. Xia is with Department of Computer and Information Science and Engineering,
University of Florida, Gainesville, FL 32611, USA.  E-mail: (yx1@cise.ufl.edu)}}

\maketitle

\begin{abstract}
We study a difficult problem of how to
schedule complex workflows with precedence constraints under a limited budget in the cloud environment.
We first formulate the scheduling problem as an integer programming problem,
which can be optimized and used as the baseline of performance.
We then consider the traditional approach of scheduling jobs in a prioritized order based
on the upward-rank of each job.
For those jobs with no precedence constraints among themselves,
the plain upward-rank priority scheme assigns priorities in an arbitrary way.
We propose a job prioritization scheme that uses Markovian chain stationary probabilities
as a measure of importance of jobs.
The scheme keeps the precedence order for the jobs that have precedence
constraints between each other, and assigns priorities according to the jobs' importance for the jobs without
precedence constraints.
We finally design a uniform spare budget splitting strategy, which splits the spare budget uniformly across all the jobs.
We test our algorithms on a variety of workflows, including FFT, Gaussian elimination, typical scientific workflows,
randomly generated workflows and workflows from an in-production cluster of an online streaming service company.
We compare our algorithms with the-state-of-art algorithms.
The empirical results show that the uniform spare budget splitting scheme outperforms the splitting scheme in
proportion to extra demand in average for most cases, and the Markovian based prioritization further improves the workflow makespan.
\end{abstract}


%
\IEEEpeerreviewmaketitle

\section{Introduction}
\label{sec:introduction}

\IEEEPARstart{T}{here} is an increasing trend to use the cloud for complex workflows, such as scientific computing workflows
and big-data analytics \cite{Juve2012} \cite{Wang2014} \cite{Rodriguez2014}.
The customers submit their workflow processing requests together with their budget to the cloud.
The workflow management system in the cloud assigns the processing requests to appropriate virtual machines (VM) by
jointly considering the requests, the VM capability and the budget. Hopefully, the customers
service level agreement will be met and the objective of the cloud provider will be optimized.
However, the current workflow management systems are
inadequate for scheduling complex workflows with diverse requirements and heterogeneous virtual machines.
This has resulted in long processing latency, wasted cloud resources, and poor return on investment.

This paper investigates a workflow scheduling problem in the cloud with budget constraints.
More specifically, a set of workflows 
is to be placed in the cloud. Each workflow has multiple
computation jobs, with precedence constraints among themselves.
For each workflow, we can use a directed acyclic graph (DAG)
to represent the precedence constraints of the jobs. A job has an execution time,
which depends on where the job is placed and how much computing resources are allocated to it.
A job has a minimum computation resource requirement, including
the CPU power and the memory requirements. The jobs are placed on a limited set of VMs.
The customer is charged only for the period when a VM is used, i.e., on the pay-as-you-go basis.
This describes the use cases of on-demand VMs in Amazon EC2.
With respect to the jobs, the decision problem we consider in this paper
is to decide where and when to place each job, i.e., which
VM will execute each job and when the execution starts. The precedence constraints and the
budget constraints must be satisfied.
Furthermore, all the resource capacity constraints at the placement targets must also be respected.
The optimization objective is to minimize the processing time of the set of workflows, i.e., the {\em makespan}
of the workflows.

Scheduling of a workflow represented by a directed task graph is a well-known NP-complete problem
in general \cite{LENSTRA1977343} \cite{Schulz1996}.
The precedence constraints among jobs make the scheduling hard and many efforts have been made to find efficient heuristics in the area of parallel computing and grid computing.
Topcuoglu et al. proposed the upward-rank based heuristic proposed in \cite{Topcuoglu02} to tackle the precedence constraints. In the upward-rank based approached, each job computes its accumulated processing time from the exit
job upward to itself along the critical (i.e., the longest) path as the upward rank. Jobs are then scheduled in
the non-increasing order of their ranks.
For the jobs with precedence constraints, the upward-rank based scheme assigns the priorities in a
reasonable way; but, for those jobs with no precedence constraints among themselves,
the upward-rank priority scheme assigns priorities in an arbitrary fashion.
In this work, we propose to assign priorities for those unrelated jobs considering jobs' importance in
the global DAG topology. We construct a random walk on the (extended) workflow DAG, and
apply the random walk stationary distribution probabilities as jobs' importance (i.e., weights).
The rationale is that the stationary probabilities are computed recursively across the global topology and
carry the global information of all states (jobs) propagated back to each state (job), and therefore
the resulted stationary probabilities reflect the jobs' importance in the global topology.
The other issue is that in parallel computing and grid computing, workflow computing often aims to minimize the makespan without considering the cost of computing facility. In the era of the cloud, the leasing cost
of the cloud facility brings a new challenge in scheduling DAG-based workflows in the cloud.
Since jobs are scheduled in a prioritized order and often greedily, how the budget is split and reserved
for each job remains a heuristic. In this work, we propose to reserve the minimum required budget for each job,
and assign the spare budget uniformly across the jobs.


We summarize the contributions of our work.
\beit
\item We formulate an integer programming model of the DAG-based workflow scheduling problem with
    budget constraints. The model can be evaluated by integer programming solvers such as {\em Gurobi}
    \cite{Gurobi} and the solution can be used as the performance baseline of different heuristics.
\item We propose a weighted upward-rank priority scheme that assigns the scheduling priorities to the jobs. It leads to improved performance in average when compared with the plain upward-rank priority scheme in \cite{Topcuoglu02}. The weights in our scheme are the stationary probabilities of a random walk on the workflow digraphs.
\item We assign the spare budget uniformly across all the jobs. The empirical results show that for most cases,
    the uniform spare-budget-splitting scheme outperforms the scheme of splitting budget in proportion to extra demand in average.
\eeit

The remaining of the paper is organized as follows. In Section \ref{sec:related}, we discuss more related works.
In Section \ref{sec:model}, we formulate the workflow scheduling problem as an integer programming problem.
We describe the weighted upward-rank priority scheme based on a random walk and the uniform spare budget splitting heuristic in Section \ref{sec:algorithm}. We evaluate the
heuristic on empirical test cases in Section \ref{sec:experiments}. Finally, we draw the conclusion in
Section \ref{sec:conclusion}.

\section{Related Works}
\label{sec:related}
DAG-based workflow scheduling has been extensively studied in the literature of parallel computing and grid computing. In the survey paper \cite{Kwok99}, the authors summarized on a wide spectrum of algorithms on DAG-based workflow scheduling in a multi-processor environment, including branch-and-bound, integer-programming, searching, randomization, and genetic algorithms. Topcuoglu et al. proposed the Heterogeneous Earliest-Finish-Time (HEFT) algorithm in \cite{Topcuoglu02}. The HEFT algorithm first computes the upward-rank of each task
by traversing the task graph; it then sorts tasks non-increasingly based on the upward-rank values, and
assigns the tasks in the sorted list to the available fastest processor. Upward-rank based task
prioritization achieves good performance and becomes an important solution in solving DAG-based workflow
scheduling.
Daoud and Kharma studied a similar problem in \cite{Daoud2008} and designed the longest dynamic critical path algorithm (LDCP).
The LDCP algorithm introduces a DAG for each processor, named DAGP, with the sizes of all the tasks set to the
computation costs of on each specified processor. It computes the upward-rank of each task within a DAGP to gain
more precise task priorities. A task with the highest upward-rank among all DAGPs is assigned with a priority to
the proper processor and all DAGPs will be updated after the assignment. The tie is broken by choosing the task
with the largest number of outgoing edges. The LDCP has better scheduling performance than HEFT, but with higher complexity.
The work in \cite{Xiang2017} studies the problem of minimizing the execution time of a workflow in heterogeneous environments and designs an ant-colony based heuristic algorithm. The heuristic generates task sequences considering both the forward and backward (i.e., global) dependency of tasks, where the forward dependency is defined as the number of predecessors, and the backward dependency is defined as the number of successors, respectively. The algorithm searches the suitable machine with a greedy minimum strategy in each round of searching. The work in \cite{Xiang2017} aligns with our opinion that not only jobs on the critical path but also other jobs should be accounted when we compute the scheduling priority.


As more and more workflows are moved to the cloud, scheduling DAG-based workflows faces a new challenge
of scheduling tasks under budget constraints.
Recently, several studies
have worked on the budget-constrained workflow makespan minimization problem in the cloud environment
\cite{Wang2014} \cite{Shu2017} \cite{Chen2017}.
Wang and Shi \cite{Wang2014} consider a special $\kappa$-stage MapReduce-like workflow where each stage consists of a batch of concurrent jobs.
Their approach is to first greedily allocate budget to the slowest job of each stage across all the stages, hoping to minimize the execution time of each stage. It then gradually refines the budget allocation across the stages and
schedules the concurrent jobs of each stage based on the budget.
Shu and Wu \cite{Shu2017} study a workflow mapping problem to minimize workflow makespan under a
budget constraint in public clouds. The work assumes that a job consists of homogeneous tasks and there is an
unlimited number of VMs in the cloud.
It pre-computes the most expensive schedule and the cheapest schedule based on the
concept of the critical path, and applies the binary search to find an approximate solution.
The work in \cite{Chen2017} considers a budget-constrained workflow scheduling heuristic in a heterogenous
cloud environment. The heuristic algorithm schedules the task in a prioritized order based on the upward-rank of each task \cite{Topcuoglu02}. The main idea of the algorithm is that it splits and reserves the budget to each individual task. It first assigns each task the minimum budget equal to the cost of using the cheapest VM; then, the
remaining budget is split so that each task gets an additional share in proportion to the cost difference between using the cheapest VM and using the most expensive VM.
Hence, by reserving the minimum budget to each task, the algorithm
guarantees to find a feasible solution. By splitting the extra budget in proportion to each task's extra cost
demand, the heuristic reserves more spare budget for the tasks with lower priorities.
These jobs will enjoy more flexibility in selecting better VMs.
Sakellariou et al. considered the facility cost in a grid environment \cite{Sakellariou2007}.
It proposes two approaches to find a minimum makespan solution with budget constraint, {\em LOSS} and
{\em GAIN}, respectively. The {\em LOSS} approach starts with the scheduling solution achieved by the HEFT algorithm, and
keeps swapping task to cheaper machines until the budget constraint is satisfied. The {\em GAIN} approach
starts with a solution with the cheapest cost, and keeps swapping tasks to faster machines whenever there is
available budget.
The work in \cite{Zheng2013} extends the HEFT algorithm in \cite{Topcuoglu02} and proposes a  Budget-constrained HEFT algorithm (BHEFT). The BHEFT algorithm assigns scheduling priorities based on the upward
rank. It splits the budget to each task based on its average cost over difference resources; if there is
additional spare budget, the spare budget will be assigned to each task in proportion to its demand. With the budget for each individual task, the BHEFT algorithm always assigns the affordable fastest resource to a task. Arabnejad and Barbosa worked on a similar DAG scheduling problem in \cite{Arabnejad2014}
and proposed the HBCS algorithm.
The task prioritization is also based on the upward-rank. The HBCS algorithm
computes a {\em worthiness} indicator which
jointly considers the cost, the remaining budget and the speed of each processor and assigns a task to the processor with the highest {\em worthiness}.

Some studies consider the min-cost workflow scheduling problem under the processing deadline constraint.
Abrishami et al. proposed the IaaS cloud partial critical paths algorithm (IC-PCP algorithm)
in \cite{ABRISHAMI2013} to minimize the execution cost of the workflow under a deadline constraint.
The key idea is the critical parent and partial critical paths(PCPs). The critical parent of a task is its unassigned parent that has the latest finish time. The PCP consists of a task and its critical parents.
The algorithm schedules tasks in a PCP as a pack, and assigns it to the cheapest VM which can meet the
sub-deadline of the PCP.
Sahni and Vidyarthi proposed the just-in-time (JIT-C) algorithm in a follow-up
work of the IC-PCP \cite{Sahni2018}. It first checks the feasibility of the customer's deadline requirement.
With a feasible deadline, the algorithm starts from the entry tasks and steps into a monitoring control loop.
Within each control loop, it identifies the tasks whose parent tasks have been scheduled and are running, and assigns each of these tasks to the cheapest VM satisfying its sub-deadline requirement.

Regarding the scheduling of multiple workflows, several different scheduling strategies were proposed.
The work in \cite{Zhao2006} focuses on how to schedule mutiple workflows onto a set of heterogeneous resources and minimize the makespan.
It proposes four policies to create a composite DAG, including common entry and common exit node, level-based ordering, alternating DAGs, and ranking-based composition.
It define a \emph{slowdown} metric as the ratio of the finish time achieved when a workflow is scheduled individually and the finish time achieved when the workflow is scheduled together with other workflows.
It aims to achieve fairness across workflows by minimizing the largest slowdown value when scheduling jobs.
The work in \cite{Xie2017} uses a heterogeneous priority rank value that includes the out-degree of a task as
a weight in the evaluation of task priorities.
It further proposes three scheduling strategies across multiple workflows including round-robin, priority-based, and trade off between round-robin and priority.
Rodriguez and Buyyawe \cite{Rodriguez2018} proposed an elastic
resource provisioning and scheduling algorithm for multiple
workflows, which aims to minimize the overall cost of leasing resources while meeting the independent
deadline constraint of workflows.

Wang and Xia explored using mixed integer programming (MIP) to formulate and solve complex workflow scheduling problems as building blocks of large-scale scheduling problems \cite{Wang2017}. The scheduling problems considered in \cite{Wang2017} are minimization of the cost under the deadline constraint.
Meena et al. \cite{Meena2016} aimed at finding schedules to minimize the execution cost while meeting the deadline in cloud computing environment. They employed a PerVar parameter to record the variation of performance of VMs and proposed a Cost Effective Genetic Algorithm (CEGA) to generate schedules.
Li et al. \cite{Li2018} focused on a similar work of \cite{Meena2016} and captured dynamic performance fluctuations of VMs by a time-series-based approach. With the VM performance forecast information, they designed a genetic algorithm that fulfills the Service-Level-Agreement.
The work in \cite{Zhou2016} develops a scheduling system to minimize the expected monetary cost given the user-specified probabilistic deadline guarantees in IaaS clouds. It focuses on dealing with the price and performance dynamics in clouds and does not assume precedence constraints in workflows.
Zheng et al. \cite{Zheng2016} studied the problem of improving utility of cloud computing by allowing partial
execution of jobs. The workflows in clouds consist of parallel homogeneous preemptable tasks without
precedence constraints. The work proposes efficient online multi-resource allocation algorithms.
Champati and Liang considered the job-machine assignment problem in the setting where
jobs have placement constraints, and machines are heterogeneous \cite{Champati2017},
and there is no precedence constraints either. They developed an efficient algorithm to
minimize the sum-cost.

\section{Problem Formulation}
\label{sec:model}

In this section, we describe the cloud system and the problem formulation.
The formulation here overlaps with the one in \cite{Wang2017}.
Assume there is a set of cloud computing workflows denoted by $\cal W$ = \{1, 2, ... , $W$\}.
For each workflow $w \in W$, it contains one or more jobs.
The total pool of jobs is denoted by $\cal J$ = \{1, 2, ... , $J$\}.
Each job $j \in \cal J$ can only belong to one workflow $w \in \cal W$.
Let ${\cal J}_w$ denote the set of jobs belonging to workflow $w$.
For job $j$, the minimum CPU requirement of job $j$ is denoted by $c_j$,
and the minimum memory requirement of job $j$ is denoted by $m_j$.
In a workflow, a job can depend on other jobs, i.e., a job cannot start until some other jobs
finish execution.
The job dependency is usually captured by a workflow DAG.
Each job in the workflow is a node in the graph and the dependency relations are denoted by directed edges between two nodes.
It is more convenient for us to represent the job dependency DAG as a matrix $L = (l_{ij})$, $\forall i,j \in \cal J$.
If job $i$ depends on job $j$, we set $l_{ij} = 1$; $l_{ij} = 0$ means that job $i$ does not depend on job $j$.
If $l_{ij} = 1$, then the start time of job $i$ should be no earlier than the finish time of job $j$, which is a precedence constraint.

For the cloud system resource, we consider a set of virtual machines (VMs) $\cal V$ = {1, 2, ... , $V$},
possibly of different types and capabilities.
Let $C_k$ represent the number of vCPUs of VM $k$,
and $M_k$ represent the amount of memory of VM $k$.
We assume a discrete time model, where time is divided into a sequence of time slots $1, 2, ... ,T$,
for instance, $5$ minutes per time slot.
At any time slot $t$, there can be at most one job allocated to any VM.
We also assume non-preemptive scheduling of jobs.
Let us characterize the amount of computation of job $j$ in terms of vCPU-time-slots, denote it by $h_j$.
Therefore when job $j$ runs on VM $k$, the running time of job $j$, $R_{jk}$, can be
computed as $R_{jk}$ = $h_j$/$C_k$, which is measured in number of time slots.
We consider the popular pay-as-you-go cloud computing that charges based on the operating time of VMs.
Suppose after running VM $k$ for a unit time, the user will be charged a cost of $p_k$.
Suppose all the workflows in question belong to the same user, which has a total budget of $D$.
We consider the problem of minimizing the finish time of all the workflows, i.e., the makespan, subject to the budget constraint and various other constraints. More specifically, for each job, we decide the VM and the starting time slot to which the job is assigned. The goal is that the overall VM leasing cost is within the budget $D$ and the makespan of all the workflows is minimized.

Next, we specify the various constraints.
Let us denote the job-VM assignment decision by the binary variables $x_{jk}^t$. We set $x_{jk}^t = 1$ if and only if job $j$ is assigned to VM $k$ and it starts at time slot $t$.
For each job $j$, only one of the $x_{jk}^t$ is equal to $1$.
\begin{align}
\label{eq:constaint_1}
\sum_{k \in \cal V} \sum_{t \in \cal T} x_{jk}^t = 1, \forall j \in \cal J.
\end{align}
When we choose the appropriate VM for job $j$, job $j$' s minimum resource requirement must be satisfied.
\begin{align}
\sum_{k \in \cal V} \sum_{t \in \cal T} C_k x_{jk}^t \geq c_j, \forall j \in \cal J. \label{eq:cpu}\\
\sum_{k \in \cal V} \sum_{t \in \cal T} M_k x_{jk}^t \geq m_j, \forall j \in \cal J. \label{eq:memory}
\end{align}

Let us discuss the precedence constraint. We note that the precedence constraint is
active only if $l_{ij} = 1$. The start time of job $i$ can be defined as
$\sum_{k \in \cal V} \sum_{t \in \cal T} t x_{ik}^t$. The finish time of job $j$
can be described as $\sum_{k \in \cal V} \sum_{t \in \cal T} ( t+R_{jk} ) x_{jk}^t$.
The precedence constraint says that if job $i$ depends on job $j$, then job $i$ cannot
start earlier than the finish time of job $j$.
\begin{align}
(\sum_{k \in \cal V} \sum_{t \in \cal T} t x_{ik}^t
-\sum_{k \in \cal V} \sum_{t \in \cal T} ( t+R_{jk} ) x_{jk}^t)l_{ij} \geq 0, \forall i,j \in \cal J.
\label{eq:precedence}
\end{align}

There is one additional constraint that at most one job runs on a VM at any time.
\begin{align}
\sum_{i \in \cal J} \sum_{r=\max(0,t-R_{ik}+1)}^t x_{ik}^r \leq 1, \forall k \in \mathcal{V}, t \in \mathcal{T}.
\label{eq:constraint_2}
\end{align}
We explain the constraint (\ref{eq:constraint_2}) in more details.
If job $i$'s execution occupies time slot $t$ of VM $k$, then job $i$'s start time
is from the set $\{max(0, t - R_{ik} + 1), ... , t\}$.
It is equivalent to saying that
$\sum_{r=\max(0,t-R_{ik}+1)}^t x_{ik}^r = 1$ for job $i$.
According to the non-preemptive requirement, at any time slot $t$, for any VM $k$,
there is at most one job that can start execution at time $t$. Therefore,
we have (\ref{eq:constraint_2}).
We show that (\ref{eq:constraint_2}) is sufficient to guarantee the existence of an non-preemptive scheduling.
Suppose for job $j$, $x_{jk}^s = 1$ for some time slot $s$ and some VM $k$.
For each time slot $t$ from $s$ to $s + R_{jk} - 1$,
together with (\ref{eq:constraint_2}) and $x_{jk}^s = 1$, we have
\begin{align}
\sum_{i \in \cal J} \sum_{r=\max(0,t-R_{ik}+1)}^t x_{ik}^r = 1. \label{eq:sufficiency}
\end{align}
Thus, for each $i \not= j$, $x_{ik}^r = 0$ for $r \in {\max(0, t - R_{ik} + 1), \cdots , t}$.
By varying $t$ from $s$ to $s + R_{jk} - 1$, we see that job $i$ cannot start on
$\{max(0, s - R_{ik} + 1), ... , s + R_{ik} - 1\}$.
We conclude that no other jobs can interfere with job $j$'s execution.

Let the variable $d$ denote an upper bound of the finish time of all the workflows. We have
\begin{align}
\sum_{k \in \cal V}\sum_{t \in \cal T} (t+R_{jk}) x_{jk}^t \leq d, \forall j \in {\cal J}. \label{eq:d}
\end{align}
%
The budget constraint of executing the workflows can be written as:
\begin{align}
\sum_{k \in \cal V} \sum_{j \in \cal J} p_k R_{jk} \sum_{t \in \cal T} x^t_{jk} \leq D.
\label{eq:budget_pay_as_you_go}
\end{align}
The workflow scheduling problem with the pay-as-you-go pricing model can be written as follows:
\begin{align}
\text{Min-Makespan:} \nonumber \\
\min \ \ \ &  d \label{eq:objective_pay_as_you_go} \\
s.t.  \ \ \ & (\ref{eq:constaint_1}) (\ref{eq:cpu}) (\ref{eq:memory}) (\ref{eq:budget_pay_as_you_go}) (\ref{eq:precedence}) (\ref{eq:constraint_2}) (\ref{eq:d}) \nonumber \\
& x, y \ \text{binary}, d \ \text{integer}. \nonumber
\end{align}

Note that data transfer costs between jobs are not directly considered in the formulation
(\ref{eq:objective_pay_as_you_go}).
We assume that data transfer takes place in the internal network of a datacenter, and the
transfer rate is stable. Therefore the data transferring time between each pair of
jobs is a constant and can be included as a part of the job's running time $R_{jk}$ \cite{Sahni2018} \cite{Meena2016} \cite{Mao2011}.

\subsection{Solve the problem by MIP software}
The Min-Makespan problem (\ref{eq:objective_pay_as_you_go}) is a complex 
integer programming problem and is usually hard to solve. {\em Gurobi} is the state-of-art MIP
software, and is capable of solving small to medium sized problems. 
We will use {\em Gurobi} to solve some instances of the Min-Makespan problem. But, the goal is to
provide a baseline for performance comparison with the heuristic algorithm that
we will propose in Section \ref{sec:algorithm}.
\section{A Motivating Example}
\label{sec:example}

\begin{figure}[htbp]
\centering
\includegraphics[height=7.5cm, width=6.5cm]{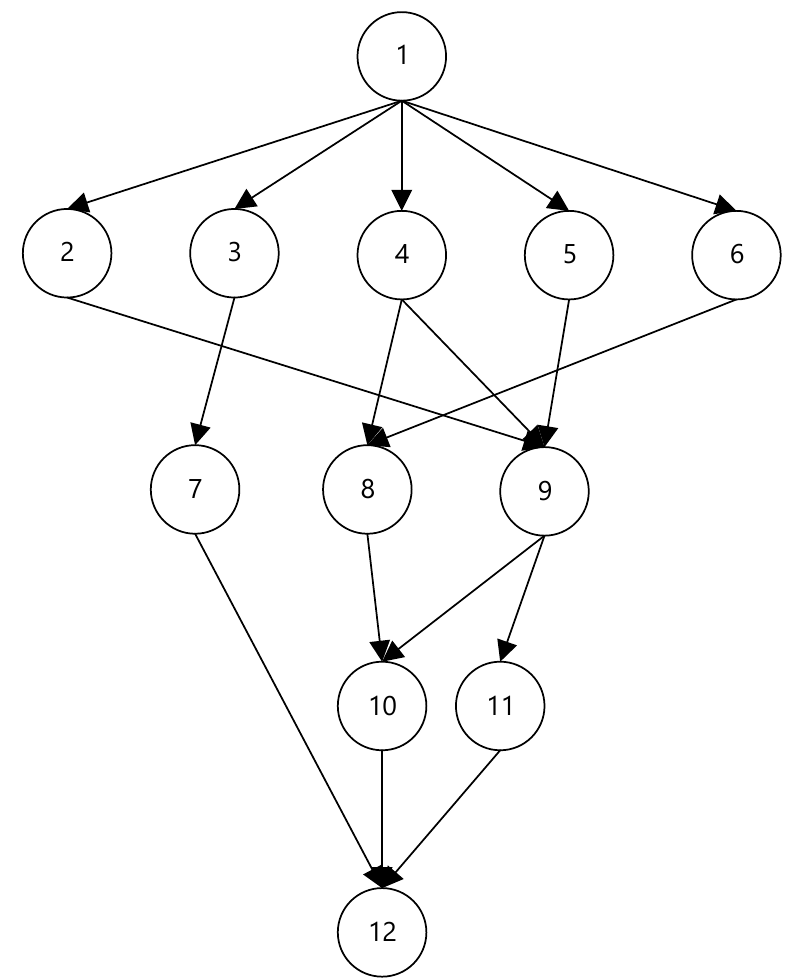}
\caption{A workflow example with $12$ jobs, job $n_1$ through job $n_{12}$.}
\label{fig:example}
\end{figure}

Consider a workflow with $12$ jobs shown in Fig. \ref{fig:example}. There are $3$ VMs and the leasing cost
of each VM is shown in Table \ref{table:VM_cost_example}. The execution time of each job on each VM is shown
in Table \ref{table:execution_time_example}.

\begin{table}
\centering
\caption{Leasing cost of the VMs in the example of Fig. \ref{fig:example}.}
\begin{tabular}{|c|c|c|c|}
\hline
 VM & $\text{VM}_1$ & $\text{VM}_2$ & $\text{VM}_3$ \\
\hline
Price   & 3	& 5	& 6\\
\hline
\end{tabular}
\label{table:VM_cost_example}
\end{table}

\begin{table}
\centering
\caption{Execution time of the jobs on each VM.}
\begin{tabular}{|c|c|c|c|}
\hline
$n_i$ & $\text{VM}_1$ & $\text{VM}_2$ & $\text{VM}_3$\\
\hline
1	&16	&14	&7\\
\hline
2	&19	&13	&16\\
\hline
3	&17	&11	&10\\  		
\hline
4	&13	&8	&15\\  			
\hline
5	&12	&13	&8\\    			
\hline
6	&13	&16	&7\\ 			
\hline
7	&6	&16	&9\\
\hline
8	&12	&11	&5\\
\hline
9	&8	&9	&11\\
\hline
10	&21	&7	&14\\
\hline
11	&12	&8	&16\\
\hline
12	&21	&7	&14\\
\hline
\end{tabular}
\label{table:execution_time_example}
\end{table}

In the well-known priority-based greedy algorithm in \cite{Topcuoglu02}, each job is assigned an upward-rank,
which is a value. The jobs are sorted in a non-increasing order according to the upward-rank, and the resulting ordered list gives the priorities to the jobs according to which the jobs are assigned to the VMs.

\subsection{Job scheduling priorities}
The upward-rank of a job $j$ is recursively defined as
\begin{align}
\bar{R}_{j} & = \frac{\sum_{k \in {\cal V}_j} R_{jk}}{|{\cal V}_j|},  \label{eq:avg_time} \\
w_{j_{\text{exit}}} & = \bar{R}_{j_{\text{exit}}}, \label{eq:uprank_exit} \\
w_j & = \bar{R}_{j} + \max_{i \in \text{succ}(j)} \{ w_i \}. \label{eq:uprank}
\end{align}
In (\ref{eq:avg_time}), ${\cal V}_j = \{k | k \in {\cal V} \text{ and } C_k \geq c_j, M_k \geq m_j\}$ is the set of the VMs that has the capacity to accept job $j$. Then, $\bar{R}_{j}$ is the average job processing time over the VM in the set ${\cal V}_j$.
The set $\text{succ}(j)$ is the set of all successor jobs of job $j$ in the workflow DAG.
The upward-rank of the exit job in the DAG, $w_{j_{\text{exit}}}$,
is defined as its average processing time.
The upward-rank of any other job, $w_j$, can be computed recursively by traversing from the exit job upward
as in (\ref{eq:uprank}). In fact, the upward-rank of a job is the aggregated upward-ranks along the critical (the longest in terms of upward-rank) path from the exit job to the current job.
In the upward-rank-based job scheduling in \cite{Topcuoglu02}, all the jobs are sorted according to the upward-rank non-increasingly;
the job with the highest upward-rank is scheduled first, and will be assigned a VM by a
separate job-VM matching algorithm, such as the HBCS algorithm in \cite{Arabnejad2014}.
We will call the priority generation scheme in \cite{Topcuoglu02} the {\em plain} upward-rank priority scheme.

In the plain upward-rank priority scheme, equation (\ref{eq:uprank}) ensures that the upward-rank of a job is higher than all its successors (including the non-immediate successors). Therefore, a job is selected with a higher priority than all its successors for VM assignment.
However, for the jobs that have no precedence
constraints among each other, the upward-rank is not a good enough indicator of a job's scheduling priority.

For instance, in Fig. \ref{fig:example}, jobs $n_3$, $n_8$ and $n_9$ do not depend on each other.
As shown in Table \ref{table:ranks_example}, the plain upward-ranks of $n_3$, $n_8$ and $n_9$ are
$38$, $38$ and $38$, respectively. Thus, the tie across jobs $n_3$, $n_8$ and $n_9$ needs to be
broken arbitrarily in scheduling. But, based on the DAG in Fig. \ref{fig:example}, jobs
$n_8$ and $n_9$ are more intricately related with other jobs in the workflow, and, to shorten the workflow makespan, it might be worthwhile to assign higher priorities to $n_8$ and $n_9$.
We will later propose a weighted upward-rank priority scheme in Section \ref{sec:algorithm}. In Table \ref{table:ranks_example}, we show the ranks and the corresponding
order of the jobs. With the weighted ranks, jobs $n_8$ and $n_9$ have higher priorities than job $n_3$, and will be scheduled earlier than $n_3$.
After generating the priority list, we apply the HBCS algorithm from \cite{Arabnejad2014} to assign each job to a VM. Tables \ref{table:example_rank} and \ref{table:example_mrank}
show the final scheduling results for the two priority generation schemes, respectively.
In Table \ref{table:example_rank}, job $n_3$ is scheduled before $n_8$ and $n_9$. Job $n_3$ occupies
the faster $\text{VM}_2$, and the final makespan is $78$.
In Table \ref{table:example_mrank}, $n_8$ and $n_9$ are assigned higher priorities because of our new priority generation scheme.
Job $n_8$ can choose the faster VM, which results in a makespan of $64$.

Hence, in assigning job scheduling priorities, we need to evaluate the importance of a job
by considering not only the jobs on its critical path but also its relationship with other jobs.

\begin{table}
\centering
\caption{Rank values and scheduling order for jobs under two different priority generation schemes strategies.}
\begin{tabular}{|c||c|c||c|c|}
\hline
Job & Upward& Scheduling& Weighted & Scheduling\\
& Rank & Order & Rank &Order \\
\hline
$n_1$ &67 & 1 &73.15 & 1 \\
\hline
$n_2$ &54 & 2 &58.29 & 2 \\
\hline
$n_3$ &38 &6 &40.72  & 8 \\
\hline
$n_4$ &50 & 3 &54.17 & 3 \\
\hline
$n_5$ &49 & 5 &53.14 &5	\\
\hline
$n_6$ &50 & 3 &54.17 &3	\\
\hline
$n_7$ &25 &11 &27.33  &11	\\
\hline
$n_8$ &38 &6  &41.81  &6	\\
\hline
$n_9$ &38 &6  &41.81  &6	\\
\hline
$n_{10}$ &28 &9  &31.23	 &9\\
\hline
$n_{11}$ &26 &10  &28.36 &10	\\
\hline
$n_{12}$ &14 & 12 &16.00 & 12\\
\hline
\end{tabular}
\label{table:ranks_example}
\end{table}

\begin{table}
\centering
\caption{Final scheduling results using HBCS with the plain upward-rank priority scheme (budget =500).}
\begin{tabular}{|c|c|c|c|c|c|c|}
\hline
$n_i$	&Budget	&Cost &Saved &Start &Finish & VM Assigned	\\
& & & Budget &Time &Time & \\
\hline
1	&100 &42 &58 &0	&7	&3	 \\
\hline
2	&115  &65 &50	&7	&20	&2		\\
\hline
4	&89 &39 &50	 	&7	&20	&1 \\
\hline
6	&89	&39 &50 &20	&33	&1	 \\
\hline
5	&86	&65 &21 &20	&33	&2	\\
\hline
3	&72	&55 &17	&33	&44	&2	\\
\hline
8	&47	&36 &11	&33	&45	&1	\\
\hline
9	&35	&24 &11	&45	&53	&1	\\
\hline
10	&46	&35 &11	&53	&60	&2	\\
\hline
11	&47	&36 &11	&53	&65	&1	\\
\hline
7	&29	&18 &11	&65	&71	&1	\\
\hline
12	&46	&35 &11	&71	&78	&2	\\
\hline
\multicolumn{7}{l}{Actual cost = 489, makespan = 78.}\\
\end{tabular}
\label{table:example_rank}
\end{table}

\begin{table}
\centering
\caption{Final scheduling results using HBCS with our weighted upward-rank priority scheme (budget =500). }
\begin{tabular}{|c|c|c|c|c|c|c|}
\hline
$n_i$	&Budget	&Cost &Saved &Start &Finish &VM Assigned	\\
& & & Budget &Time &Time & \\
\hline
1	&100 &42 &58	&0	&7	&3	\\
\hline
2	&115 &65 &50	 &7	&20	&2	\\
\hline
4	&89	&39	&50	&7	&20	&1	\\
\hline
6	&89	&42	&47	&7	&14	&3	\\
\hline
5	&83	&48	&35	&14	&22	&3	\\
\hline
8	&65	&55	&10	&20	&31	&2	\\
\hline
9	&34	&24	&10	&22	&30	&1	\\
\hline
3	&61	&55	&6	&31	&42	&2	\\
\hline
10	&41	&35	&6	&42	&49	&2	\\
\hline
11	&42	&40	&2	&49	&57	&2	\\
\hline
7	&20	&18	&2	&42	&48	&1	\\
\hline
12	&37	&35	&2	&57	&64	&2	\\
\hline
\multicolumn{7}{l}{Actual cost = 498, makespan = 64.}\\
\end{tabular}
\label{table:example_mrank}
\end{table}

\subsection{Budget splitting}
In HBCS, the spare budget is preferentially assigned to the jobs with
the higher priority. Because of the greedy nature of HBCS, the jobs with higher priorities tend to
use more expensive and faster VMs, whereas the jobs with lower priorities often do not have too many options
because the remaining balance is more limited.

From Table \ref{table:example_rank} and Table \ref{table:example_mrank},
it can be seen that the available budget for the jobs with lower priorities is very limited under
both priority generation schemes.
If we split the spare budget evenly as shown in Table \ref{table:example_uniform_budget}, more budget will be
allocated to jobs with lower priorities. These jobs will enjoy more flexibility in selecting better VMs, which results in shorter makespan, as shown in Table \ref{table:example_uniform_budget}.
The conclusion is that the spare budget should be split across the jobs more evenly.

\begin{table}
\centering
\caption{Final scheduling results using the plain upward-rank priority scheme and uniform spare budget splitting (budget =500).}
\begin{tabular}{|c|c|c|c|c|c|c|}
\hline
$n_i$	&Budget	&Cost &Saved &Start &Finish &VM Assigned	\\
& & & Budget &Time &Time & \\
\hline
1	&47	&42 &5 &0	&7	&3	\\
\hline
2	&67	&65 &2	&7	&20	&2	\\
\hline
4	&46	&39 &7	&7	&20	&1	\\
\hline
6	&50	&42 &8	&7	&14	&3 \\
\hline
5	&49	&48 &1	&14	&22	&3	 \\
\hline
3	&57	&55 &2	&20	&31	&2 \\
\hline
8	&37	&30	 &7	&22	&27	&3	\\
\hline
9	&36	&24	 &12	&22	&30	&1 \\
\hline
10	&52	&35	 &17	&31	&38	&2 \\
\hline
11	&57	&36	 &21	&30	&42	&1 \\
\hline
7	&44	&18	 &26	&42	&48	&1 \\
\hline
12	&66	&35 & 31 &48 &55 &2 \\
\hline
\multicolumn{7}{l}{Actual cost = 469 $\leq$ Budget, makespan = 55.}\\
\end{tabular}
\label{table:example_uniform_budget}
\end{table} 
\section{A Heuristic Algorithm}
\label{sec:algorithm}

\begin{figure}[htbp]
\centering
\includegraphics[width=3.5in]{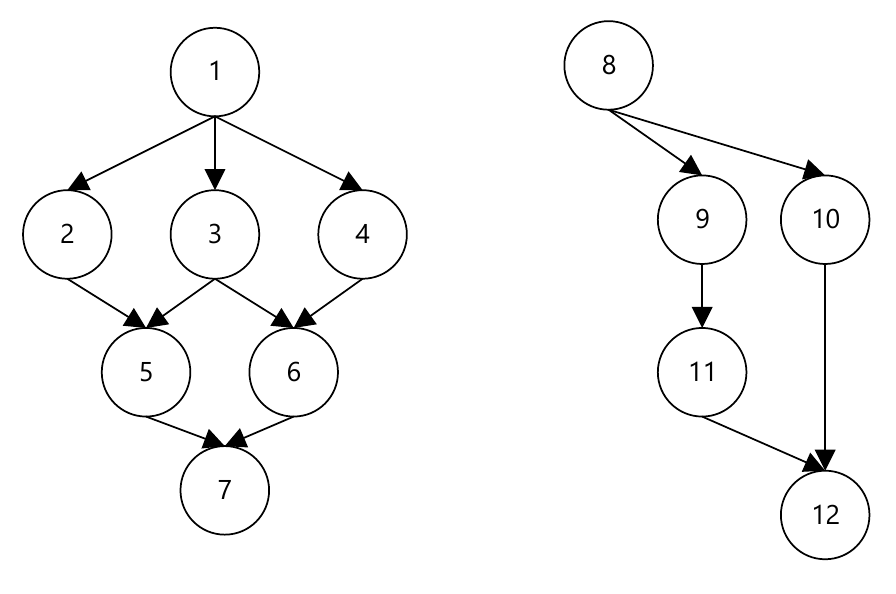}
\caption{An example of two workflows with $12$ jobs, job $n_1$ through job $n_{12}$.}
\label{fig:two_single_workflow}
\end{figure}

\begin{figure}[htbp]
\centering
\includegraphics[width=3.5in]{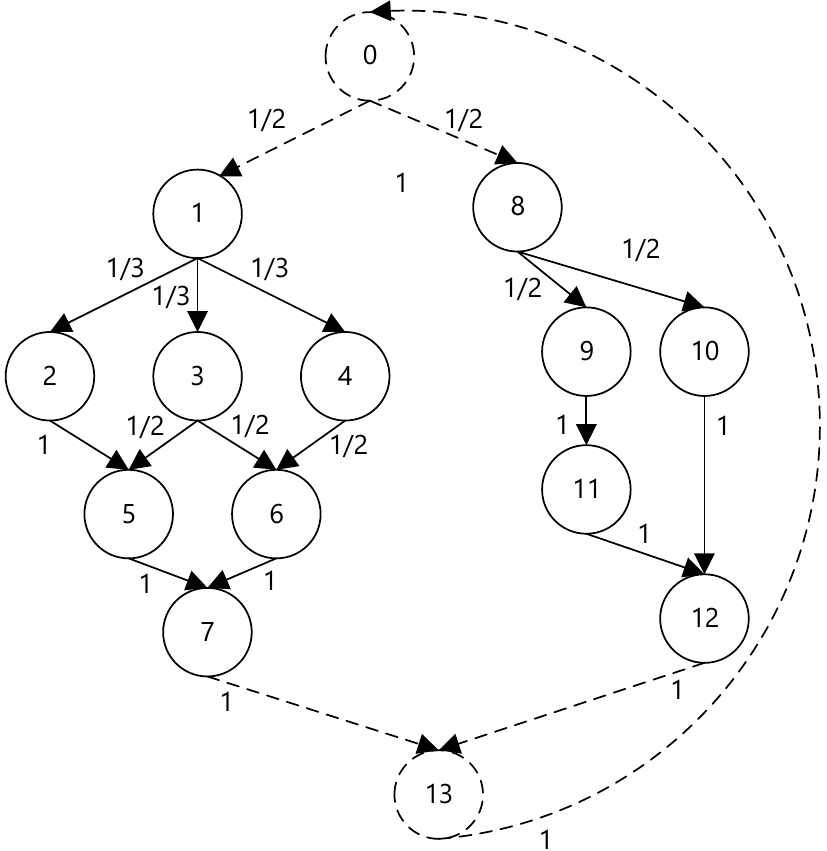}
\caption{The extended DAG by adding pesudo entry and exit nodes, job $n_0$ and $n_{13}$. The numbers around the edges are the transition probabilities of the digraph.}
\label{fig:multiple_workflow}
\end{figure}

Motivated by the example in Section \ref{sec:algorithm}, we develop a heuristic algorithm to solve the Min-Makespan problem. The algorithm has two key components. One is the weighted upward-rank priority scheme, which uses the stationary distribution of a random walk on the DAG as the weights. The other is uniform spare budget splitting.
For scheduling multiple workflows, we make an extended DAG by adding pseudo entry and exit nodes to connect multiple DAGs. The scheduling priorities of the jobs across all the workflows are computed based on the extended DAG.
In Fig. \ref{fig:two_single_workflow}, we show two typical workflow DAGs. By adding pseudo entry and exit nodes, job $n_0$ and $n_{13}$,
we have an extended DAG shown in Fig. \ref{fig:multiple_workflow}.

\subsection{Weighted upward-rank priority scheme using random walk}
According to the discussion in Section \ref{sec:example}, when we compute a job's scheduling priority, it needs to consider both the jobs on the critical path and the other jobs as well.
We follow the upward-rank based priority scheme originally proposed in \cite{Topcuoglu02}.
We propose to construct a random walk on the (extended) workflow DAG, and extend
the plain scheme by applying the random walk stationary distribution probabilities as weights to the plain ranks.
More specifically, for each job $j$,
the plain upward rank represents the accumulated processing time of successors on its critical path, and
its weight (i.e., the stationary probability $\pi_j$) represents the importance of job $j$ in the global
DAG topology.
The rationale behind is that if a job is more complicated related with other jobs in the topology, the job is
more important and deserves a higher priority as discussed in Section \ref{sec:example}.
The stationary probability vector $\pi$ of the random walk on the workflow
DAG can be interpreted as the recurrence probability of each state in the limited distribution. Generally,
if a state $j$'s stationary probability $\pi_j$ is higher than other states,
it implies that the system state prefers to transit from other states to state $j$ and state $j$ is more
important. Hence the vector $\pi$ is a good indicator of the importance of jobs and can be used as weights
of the plain upward-rank.

We describe the detailed procedure of construction the random walk.
Because the DAG is acyclic, we add directed edges to the DAG from each exit node to each entry node.
In the new graph,
the set of successors of any node $j$ is not empty.
The random walk is on this digraph. Let the transition probability from job (state) $j$ to job (state) $i$ be denoted by $p_{ji}$. We set
\begin{align}
\label{eq:transit_prob}
p_{ji} = \frac{1}{|succ(j)|}, \forall i, j \text{ where } l_{ij} = 1.
\end{align}
Thus, from state $j$, the random walk will visit its immediate successors with equal probabilities.
Note that, if job $i$ does not depend on job $j$, then $p_{ji} = 0$. We show the transition probabilities
of an example DAG in Fig. \ref{fig:multiple_workflow}.

Let $\pi_j$ denote the stationary probability for state $j$. The stationary probabilities can be computed
by solving the following equations.
\begin{align}
& \sum_{j \in {\cal J}} \pi_j = 1, \label{eq:stationary_distr}\\
& \sum_{i} p_{ij} \pi_i = \pi_j, \forall j. \label{eq:stationary_distr_2}
\end{align}

We use the stationary probability $\pi_j$ as a measure of importance of job $j$.
The weighted upward-ranks
are defined recursively as follows.
\begin{align}
w_{j_{\text{exit}}} & = \bar{R}_{j_{\text{exit}}} \pi_{j_{\text{exit}}},
\label{eq:uprank_exit_markov} \\
w_j & = \bar{R}_{j} \pi_j + \max_{i \in \text{succ}(j)} \{ w_i \}. \label{eq:uprank_markov}
\end{align}

\subsection{Uniform spare budget splitting}
After the jobs' scheduling priorities are determined, we need to split the budget across the jobs.
In order to guarantee that each job can rent a VM,
a job $j$ needs to receive a minimum budget, denoted by $D^{\min}_j$, given by
\begin{align}
\label{eq:min_job_budget}
D^{\min}_j = \min_{k \in {\cal V}_j} \{ p_k R_{jk} \}.
\end{align}
Thus, a feasible budget $D$ should be no less than the aggregate minimum budget of each job, i.e.,
\begin{align}
\label{eq:min_workflow_budget}
D \geq \sum_{j \in \cal J} D^{\min}_j.
\end{align}
For the spare budget $D - \sum_{j \in \cal J} D^{\min}_j$, we propose to split it evenly across the jobs.
Hence, the reserved budget of each job $j$ is computed as
\begin{align}
\label{eq:reserve_job_budget}
D^{\text{reserve}}_j = D^{\min}_j + \frac{D - \sum_{j \in \cal J} D^{\min}_j}{|{\cal J}|}.
\end{align}


We summarize the overall scheduling algorithm in Algorithm \ref{algo:scheduling}.
Note that in Step $1(e)$, we can also use the plain upward-rank priority scheme. The resulting algorithm is still a new algorithm, compared with the algorithm in \cite{Topcuoglu02}, because of the new way of splitting the spare budget - uniform splitting.

\begin{algorithm}
\caption{Multiple workflow scheduling with the weigthed upward-rank priority scheme and uniform spare budget splitting}
\label{algo:scheduling}
\beit
\item Initialize:
    \beit
    \item   Step $1(a)$: Add a pseudo entry node and a pseudo exit node with computation cost
        $h_{\text{entry}} = h_{\text{exit}} = 0$.
    \item   Step $1(b)$: Make the entry nodes of all workflows immediate successors of the pseudo entry node,
            and the exit nodes of all workflows immediate ancestors of the pseudo exit node.
    \item   Step $1(c)$: Assign the Markov chain transition probabilities according to equations
            (\ref{eq:transit_prob}).
    \item   Step $1(d)$: Compute Markov chain stationary probabilities based on equations (\ref{eq:stationary_distr}) and (\ref{eq:stationary_distr_2}).
    \item   Step $1(e)$: Compute the weighted upward-ranks as in equations (\ref{eq:uprank_exit_markov}) and (\ref{eq:uprank_markov}).
    \item   Step $1(f)$: Sort all jobs non-increasingly according to $w_j$.
    \item   Step $1(g)$: Compute the reserved budget for each job $j$ according to equation (\ref{eq:reserve_job_budget}).
    \item   Step $1(h)$: Set the remaining balance $D_{\text{remain}}$ to be $0$.
    \eeit
\item Step $2$: Remove the job $j$ with the highest $w_j$ from the sorted list.
\item Step $3$: Select a VM $k$ from the VM set ${\cal V}_j$, where VM $k$ is the fastest VM for job $j$ within the budget limit as the follows,
    \begin{align}
        \min \ \ \ & R_{jk}  \label{eq:select_VM} \\
        s.t. \ \ \ & p_k R_{jk} \leq D_j^{\text{reserve}} + D_{{\text{remain}}} \nonumber \\
              & k \in {\cal V}_j, \nonumber
    \end{align}
    and assign job $j$ to VM $k$.
    If problem (\ref{eq:select_VM}) has multiple solutions, we will choose the cheapest VM. Any further tie will be broken arbitrarily.
\item Step $4$: Recompute the remaining balance $D_{\text{remain}}$ as
            \begin{align}
            D_{\text{remain}} = D_{\text{remain}} + (D_j^{\text{reserve}} - p_k R_{jk}). \label{eq:remain_balance}
            \end{align}
\item Step $5$: If the job list is empty, exit; else, go to Step $2$.
\eeit
\end{algorithm} 
\section{Experiments}
\label{sec:experiments}

\begin{table}
\centering
\caption{VM types in the experiments.}
\begin{tabular}{|c|c|c|c|}
\hline
VM Type & vCPU & Memory(GiB) & Price(\$/hour)\\
\hline
t2.micro & 1 & 1 & 0.0116\\
\hline
t2.medium & 2 & 4 & 0.0464\\
\hline
m5.xlarge & 4 & 16 & 0.192\\  		
\hline
m5.2xlarge & 8 & 32 & 0.384\\  			
\hline
m5.4xlarge & 16 & 64 & 0.768\\    			
\hline
m5.12xlarge & 48 & 192 & 2.304\\ 			
\hline
c5.large	&2	&4	  &0.085\\
\hline
c5.xlarge	&4	&8	  &0.17\\
\hline
c5.2xlarge	&8	&16	  &0.34\\
\hline
c5.4xlarge	&16	&32	  &0.68\\
\hline
c5.9xlarge	&36	&72	  &1.53\\
\hline
c5.18xlarge	&72	&144  &3.06\\
\hline
r4.large	&2	&15.25	&0.133\\
\hline
r4.xlarge	&4	&30.5	&0.266\\
\hline
r4.2xlarge	&8	&61	    &0.532\\
\hline
r4.4xlarge	&16	&122	&1.064\\
\hline
r4.8xlarge	&32	&244	&2.128\\
\hline
i3.xlarge	&4	&30.5	&0.312\\
\hline
i3.2xlarge	&8	&61	    &0.624\\
\hline
i3.4xlarge	&16	&122	&1.248\\
\hline
i3.8xlarge	&32	&244	&2.496\\
\hline
g3.4xlarge	&16	&122	&1.14\\
\hline
g3.8xlarge	&32	&244	&2.28\\
\hline
\end{tabular}
\label{table:VM_type}
\end{table}

In this section, we present the comparative evaluation results of our algorithms, the algorithms
of MSLBL \cite{Chen2017}, HBCS \cite{Arabnejad2014} and BHEFT \cite{Zheng2013}, and the optimal baseline solution generated by Gurobi. In Table \ref{table:shorthands}, we list the shorthands for these algorithms,
which will be used throughout this section.
We first describe a single workflow scenario, where various experimental cases and algorithms are tested
and results are reported. Then, we move to a multiple workflow scenario, where we compare our algorithm with
random and round-robin priority generation schemes.
In the experiments, we use a broad range of workloads, including workflows from real applications and randomly generated workflows.

\begin{table}
\centering
\caption{Shorthands for different algorithms.}
\begin{tabular}{|c|p{6cm}|}
\hline
BAVE	& Algorithm \ref{algo:scheduling} with the plain upward-rank priority scheme \\
\hline
BAVE\_M	& Algorithm \ref{algo:scheduling} with the weighted upward-rank priority scheme \\
\hline
MSLBL	& Algorithm in \cite{Chen2017} with the plain upward-rank priority scheme \\
\hline
MSLBL\_M	& Algorithm in \cite{Chen2017} with the weighted upward-rank priority scheme \\
\hline
HBCS & Algorithm in \cite{Arabnejad2014} \\
\hline
BHEFT & Algorithm in \cite{Zheng2013} \\
\hline
Gurobi & the optimal baseline solution generated by Gurobi \\
\hline
\end{tabular}
\label{table:shorthands}
\end{table}

\subsection{Workflow setup}
\label{subset:workflow_setup}
We use four types of real-world workflows including the Fast Fourier transform parallel application (FFT), Gaussian elimination parallel application \cite{Topcuoglu02}, scientific workflows, and real in-production workflows from an Internet streaming service company in China.

In generating the FFT workflows, we use a parameter $m$ to set the size of the FFT application. The number of jobs is $N = 2m - 1 + m \log_2 m$, where $m = 2k$ for some integer $k$. Furthermore, an FFT workflow enjoys a
symmetry. The aggregated execution time of the jobs on any path from the starting job to any of the exiting jobs is equal. Thus, any path in an FFT DAG is a critical path.
For the Gaussion elimination workflows, the number of jobs is set to be $N = \frac{n^2 + n - 2}{2}$, where $n$ is the number of rows of a square matrix.
We also evaluate other scientific workflows including Montage, CyberShake, Epigenomics, LIGO Inspiral Analysis and SIPHT, which are by an open source scientific workflow generator \cite{Silva2014}.

Finally, we obtained a $19$-hour-long logs of an in-production cluster from an Internet streaming service company in China. The cluster carried $2347$ workflows including MapReduce, Spark, Hive, Shell during the $19$ hours.
A workflow may contain multiple jobs, and a job may contain multiple parallel tasks.
The logs show that $82.7\%$ workflows only contain no more than $5$ jobs; the remaining $17.3\%$ workflows
contain the number of jobs ranging from $6$ to $375$, and these workflows occupy more than $60\%$ of the CPU and
memory resources. We evaluate the algorithms on $5$ typical workflows with different numbers of jobs.

%
%
%


\subsection{Other parameters}
We resort to simulation to compare the algorithms. All simulation experiments are conducted on a PC platform with an Intel Core $i7$ $2.60$ GHz CPU and $8$ GB memory. We use $23$ VM types as tabulated in Table \ref{table:VM_type}, which follow the VM setup in Amazon’s EC2 as close as we can \cite{EC2Inst}.
To see the influence of the number of available VM, we test with three different levels of VM sufficiency:
{\em Scarce}, {\em Normal} and {\em Sufficient}. In the {\em Scarce} case, the number of VMs is half of the number of jobs;
in the {\em Normal} case, the two numbers are equal; in the {\em Sufficient} case, the number of VMs is $1.5$ times of the number of jobs.
In all the three cases, $2/3$ of the VM instances are assigned to the VM types with no more than $8$
vCPUs in Table \ref{table:VM_type}; and the other $1/3$ of the VM instances are assigned to the VM
types with more than $8$ vCPUs. Finally, the number of instances of each VM type is generated randomly.

We also vary the budget as in (\ref{eq:budget_level}).
\begin{align}
D = D_{\min} + \varphi (D_{\max} - D_{\min}), \label{eq:budget_level}
\end{align}
where $D_{\min}$ is the cost of using the cheapest schedule, and $D_{\max}$ is the cost
obtained by the HEFT algorithm.
The budget level factor $\varphi \in \{0, 0.25, 0.5, 0.75, 1.0 \} $ is used to vary the budget level.

Finally, sometimes an algorithm may fail to find a feasible schedule, either because of the high complexity of the algorithm, or due to the greedy nature. When that happens, a failure is reported. We
report the algorithm success rates in the results.

\subsection{Summary of performance ranking}

We first summarize the overall performance of Gurobi, BAVE, BAVE\_M, MSLBL, MSLBL\_M, HBCS and BHEFT by
counting their ranks in terms of the obtained makespans.
For each test case, we order the algorithms in the increasing order of makespans; then we count their
ranks for each type of workflows. We also use an average ranked value (AR) proposed in \cite{Zheng2013} to evaluate
the performance of algorithms.
The value AR is defined as
\begin{align}
AR = \frac{R_1 + 2 R_2 + 3 R_3 + 4 R_4}{N_{cases}},
\end{align}
where $N_{cases}$ is the number of test cases, and $R_i$ is the count for rank $i$.
A smaller AR value of an algorithm stands for a better performance in average.
In Table \ref{table:FFT_rank} - \ref{table:iqiyi_rank}, the results of rank counting
and the associated AR values are reported.
For brevity, only the counting of the first three places are shown in tables.
For Gurobi, HBCS and BHEFT algorithms, because they sometimes fail to find feasible solutions, their AR values are not reported.
By inspecting the AR values of all workflow types, we draw the following conclusion.
\beit
\item
For FFT and randomly generated workflows, BAVE algorithms achieves the best performance
in average. The uniform extra budget splitting scheme outperforms the scheme of
splitting budget in proportion to extra demand. The weighted priority scheme
cannot further improve the makespan
when it combines with the uniform extra budget splitting scheme. For FFT workflows, the weighted scheme even
results in longer makespans in several test cases. Anyway, when we apply the weighted priority
scheme to MSLBL algorithm, the makespans are reduced.
\item
For Guassian and other scientific workflows, and the workflows obtained from the in-production cluster,
BAVE\_M algorithm achieves the best performance in average.
Both the weighted priority and uniform extra budget splitting schemes help
to improve the makespans.
\eeit
We conclude that the weighted priority scheme using random walk
and the uniform spare budget splitting strategy help
to improve the makespans in average for most of the test cases.

\begin{table}[H]
\centering
\caption{Ranking counts for the FFT workflow, 75 test cases}
\begin{tabular}{|c|c|c|c|c|}
\hline
RANK & 1 & 2 & 3 & AR \\
\hline
Gurobi	&15	&0	&0	&/ \\
\hline
BAVE	&64	&9	&2	&1.17 \\
\hline
BAVE\_M	&63	&8	&3	&1.23 \\
\hline
MSLBL	&31	&35	&8	&1.72 \\
\hline
MSLBL\_M &32 &34 &9	&1.69 \\
\hline
HBCS	&13	&9	&10	&/ \\
\hline
BHEFT	&9	&13	&3	&/ \\
\hline
\end{tabular}
\label{table:FFT_rank}
\end{table}

\begin{table}[H]
\centering
\caption{Ranking counts for the Gaussian workflow, 60 test cases}
\begin{tabular}{|c|c|c|c|c|}
\hline
RANK & 1 & 2 & 3 & AR \\
\hline
Gurobi	&15	&0	&0	&/ \\
\hline
BAVE	&35	&14	&5	&1.70 \\
\hline
BAVE\_M	&41	&16	&3	&1.36 \\
\hline
MSLBL	&15	&10	&20	&2.60 \\
\hline
MSLBL\_M &18 &14 &22 &2.26 \\
\hline
HBCS	&12	&0	&0	&/ \\
\hline
BHEFT	&3	&10	&0	&/ \\
\hline
\end{tabular}
\label{table:Guassian_rank}
\end{table}

\begin{table}[H]
\centering
\caption{Ranking counts for the CyberShake workflow, 15 test cases}
\begin{tabular}{|c|c|c|c|c|}
\hline
RANK & 1 & 2 & 3 & AR \\
\hline
BAVE	&6	&3	&4	&2.13 \\
\hline
BAVE\_M	&10	&4	&1	&1.40 \\
\hline
MSLBL	&6	&4	&5	&1.93 \\
\hline
MSLBL\_M &8	&5	&2	&1.60 \\
\hline
HBCS	&3	&0	&0	&/ \\
\hline
BHEFT	&3	&0	&0	&/ \\
\hline
\end{tabular}
\label{table:Cybershake_rank}
\end{table}

\begin{table}[H]
\centering
\caption{Ranking counts for the Epigenomics workflow, 15 test cases}
\begin{tabular}{|c|c|c|c|c|}
\hline
RANK & 1 & 2 & 3 & AR \\
\hline
BAVE	&6	&2	&4	&2.27 \\
\hline
BAVE\_M	&10	&4	&1	&1.40 \\
\hline
MSLBL	&4	&2	&6	&2.53 \\
\hline
MSLBL\_M &7	&6	&2	&1.67 \\
\hline
HBCS	&3	&0	&0	&/ \\
\hline
BHEFT	&2	&1	&0	&/ \\
\hline
\end{tabular}
\label{table:Epigenomics_rank}
\end{table}

\begin{table}[H]
\centering
\caption{Ranking counts for the Inspiral workflow, 15 test cases}
\begin{tabular}{|c|c|c|c|c|}
\hline
RANK & 1 & 2 & 3 & AR \\
\hline
BAVE	&7	&0	&3	&2.40 \\
\hline
BAVE\_M	&8	&5	&1	&1.67 \\
\hline
MSLBL	&4	&1	&6	&2.67 \\
\hline
MSLBL\_M &6	&7	&2	&1.73 \\
\hline
HBCS	&3	&0	&0	&/ \\
\hline
BHEFT	&1	&0	&0	&/ \\
\hline
\end{tabular}
\label{table:Inspiral_rank}
\end{table}

\begin{table}[H]
\centering
\caption{Ranking counts for the Montage workflow, 15 test cases}
\begin{tabular}{|c|c|c|c|c|}
\hline
RANK & 1 & 2 & 3 & AR \\
\hline
BAVE	&8	&3	&1	&1.93 \\
\hline
BAVE\_M	&11	&1	&3	&1.47 \\
\hline
MSLBL	&3	&4	&4	&2.60 \\
\hline
MSLBL\_M &5	&3	&6	&2.20 \\
\hline
HBCS	&3	&0	&0	&/ \\
\hline
BHEFT	&1	&1	&0	&/ \\
\hline
\end{tabular}
\label{table:Montage_rank}
\end{table}

\begin{table}[H]
\centering
\caption{Ranking counts for the Sipht workflow, 15 test cases}
\begin{tabular}{|c|c|c|c|c|}
\hline
RANK & 1 & 2 & 3 & AR \\
\hline
BAVE	&7	&1	&0	&2.47 \\
\hline
BAVE\_M	&6	&3	&6	&2.00 \\
\hline
MSLBL	&4	&5	&4	&2.33 \\
\hline
MSLBL\_M &8	&5	&1	&1.67 \\
\hline
HBCS	&3	&0	&0	&/ \\
\hline
BHEFT	&1	&1	&0	&/ \\
\hline
\end{tabular}
\label{table:Sipht_rank}
\end{table}

\begin{table}[H]
\centering
\caption{Ranking counts for the random workflow, 45 test cases}
\begin{tabular}{|c|c|c|c|c|}
\hline
RANK & 1 & 2 & 3 & AR \\
\hline
Gurobi	&15	&0	&0	&/ \\
\hline
BAVE	&32	&12	&1	&1.31 \\
\hline
BAVE\_M	&32	&12	&1	&1.31 \\
\hline
MSLBL	&11	&8	&19	&2.48 \\
\hline
MSLBL\_M &11&18	&15	&2.13 \\
\hline
HBCS	&8	&1	&0	&/ \\
\hline
BHEFT	&9	&1	&0	&/ \\
\hline
\end{tabular}
\label{table:Random_rank}
\end{table}

\begin{table}[H]
\centering
\caption{Ranking counts for the workflow obtained from an Internet company, 45 test cases}
\begin{tabular}{|c|c|c|c|c|}
\hline
RANK & 1 & 2 & 3 & AR \\
\hline
Gurobi	&15	&0	&0	&/ \\
\hline
BAVE	&20	&19	&3	&1.75 \\
\hline
BAVE\_M	&33	&12	&0	&1.26 \\
\hline
MSLBL	&10	&8	&18	&2.60 \\
\hline
MSLBL\_M &10 &13 &18 &2.35 \\
\hline
HBCS	&7	&2	&0	&/ \\
\hline
BHEFT	&0	&4	&1	&/ \\
\hline
\end{tabular}
\label{table:iqiyi_rank}
\end{table}

\begin{figure}
\begin{center}
\includegraphics[width=3.5in]{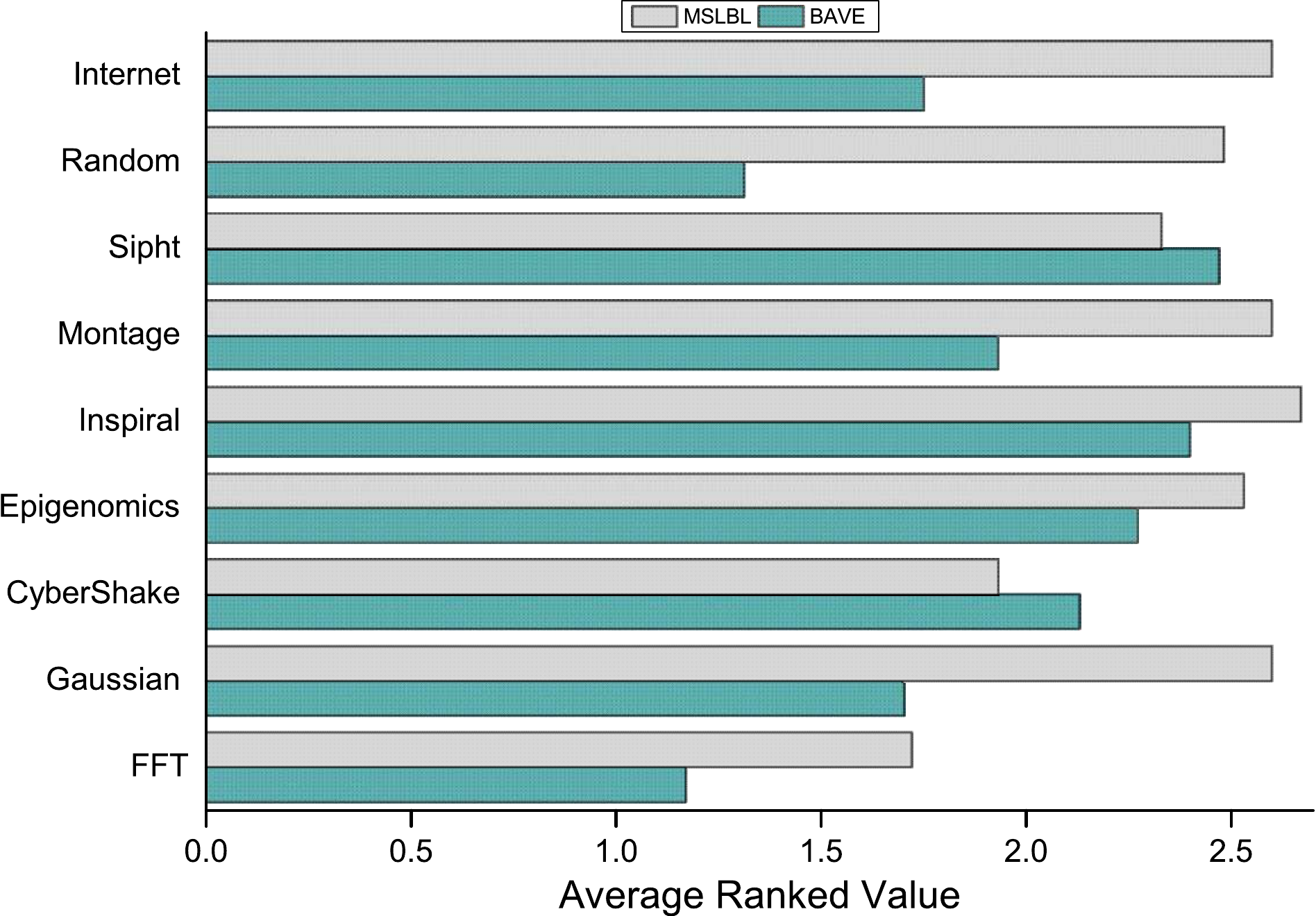}
\end{center}
\caption{Comparison of Ranking counts between BAVE and MSLBL}
\label{fig:summary1_rank}
\end{figure}

\begin{figure}
\begin{center}
\includegraphics[width=3.5in]{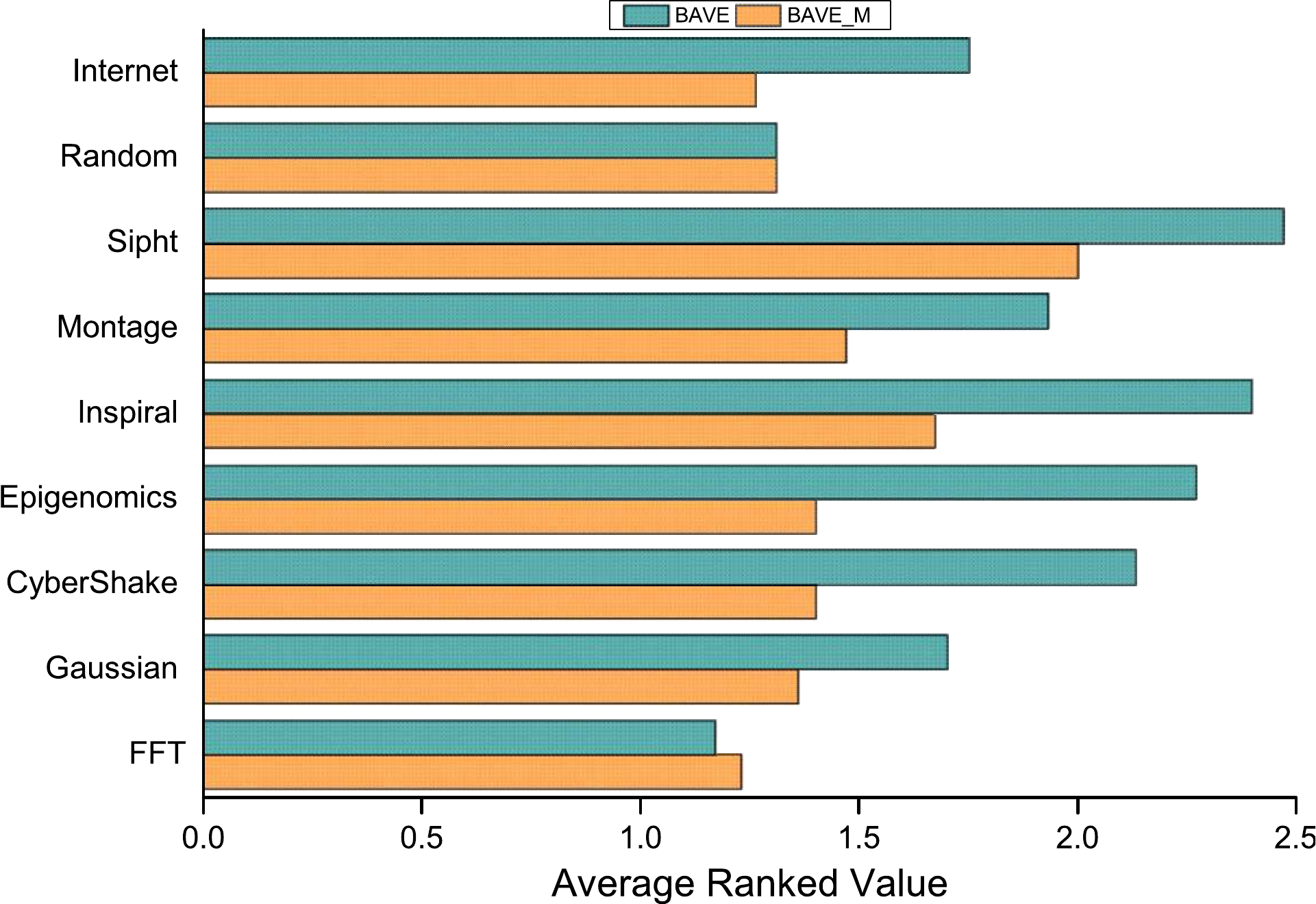}
\end{center}
\caption{Comparison of Ranking counts between BAVE and BAVE\_M}
\label{fig:summary2_rank}
\end{figure}

Finally, in order to separate out the improvement achieved by the weighted priority scheme and that by the uniform
spare budget splitting scheme, we compare the algorithm ranking results between BAVE and MSLBL, and the results between
BAVE and BAVE\_M in Fig. \ref{fig:summary1_rank} and Fig. \ref{fig:summary2_rank}, respectively.
In Fig. \ref{fig:summary1_rank}, algorithm BAVE outperforms algorithm MSLBL
on average for most workflow types except the workflow types CyberShake and Sipht. The results can be interpreted as
showing the advantage of the uniform spare budget splitting scheme. Fig. \ref{fig:summary2_rank} shows that the weighted priority scheme
improves the average performance compared with the plain priority scheme by decreasing AR for most workflow types
except the workflow types FFT and random.

\subsection{Detailed experimental results}
In this section, we plot and discuss the detailed performance of each test case.

\subsubsection{FFT}
In Fig. \ref{fig:fft_vm_nm}, we show the normalized makespans for the FFT workflows.
In each test case, the makespans obtained by different algorithms are normalized with respect to
the smallest one and are plotted to show the performance. For instance, In Fig. \ref{fig:fft_vm_nm}(a),
we show an FFT workflow with $15$ jobs. All the algorithms are tested with different levels of VM sufficiency and
budgets. Gurobi achieves the best makespan with sufficient VMs and budget level of $\varphi = 1.0$, and
hence the makespans obtained by other algorithms and settings are normalized with respect to this specific
optimal value in the plot shown in Fig. \ref{fig:fft_vm_nm}(a).

In the small experiment with $N=15$ jobs, Gurobi always achieves the best makespan which can be used as the
performance baseline. In the VM sufficiency case of {\em Scarce}, the results show that the budget level
$\varphi$ has a great impact on the resulted makespan, which
drops quickly with the increased
budget level $\varphi$. In the VM sufficiency case of {\em Normal}, the makespan improves with $\varphi$
when $\varphi$ is small. The difference of makespans between $\varphi =0.0$ and $\varphi = 1.0$ is significantly
narrowed. When there are {\em Sufficient} supplies of VMs, the makespan is further reduced when the budget is plenty. We notice that Gurobi produces solutions with slightly better makespan only in the cases of {\em Scarce}, and the case of $\varphi=0.0$ and VM sufficiency {\em Normal}.
The algorithms BAVE and BAVE\_M achieve almost the same performance as Gurobi.
The makespans produced by MSLBL and  MSLBL\_M are slightly worse in the cases of {\em Scarce}. The HBCS and BHEFT algorithms cannot always find a solution.

When the size of workflow increases to $N=95$ jobs, Gurobi fails to find a solution. There are $52,448$
binary variables in the problem formulation, which is extremely large for an integer programming problem.
In the experiments with $N=95, 223, 1151$ and $2559$ jobs, the algorithms BAVE and BAVE\_M achieve the
best makespan in almost all the test cases.
The algorithms HBCS and BHEFT cannot always find a solution.


\begin{figure}[H]
\begin{center}
\begin{minipage}{3.5in}
\begin{center}
\includegraphics[width=3.5in]{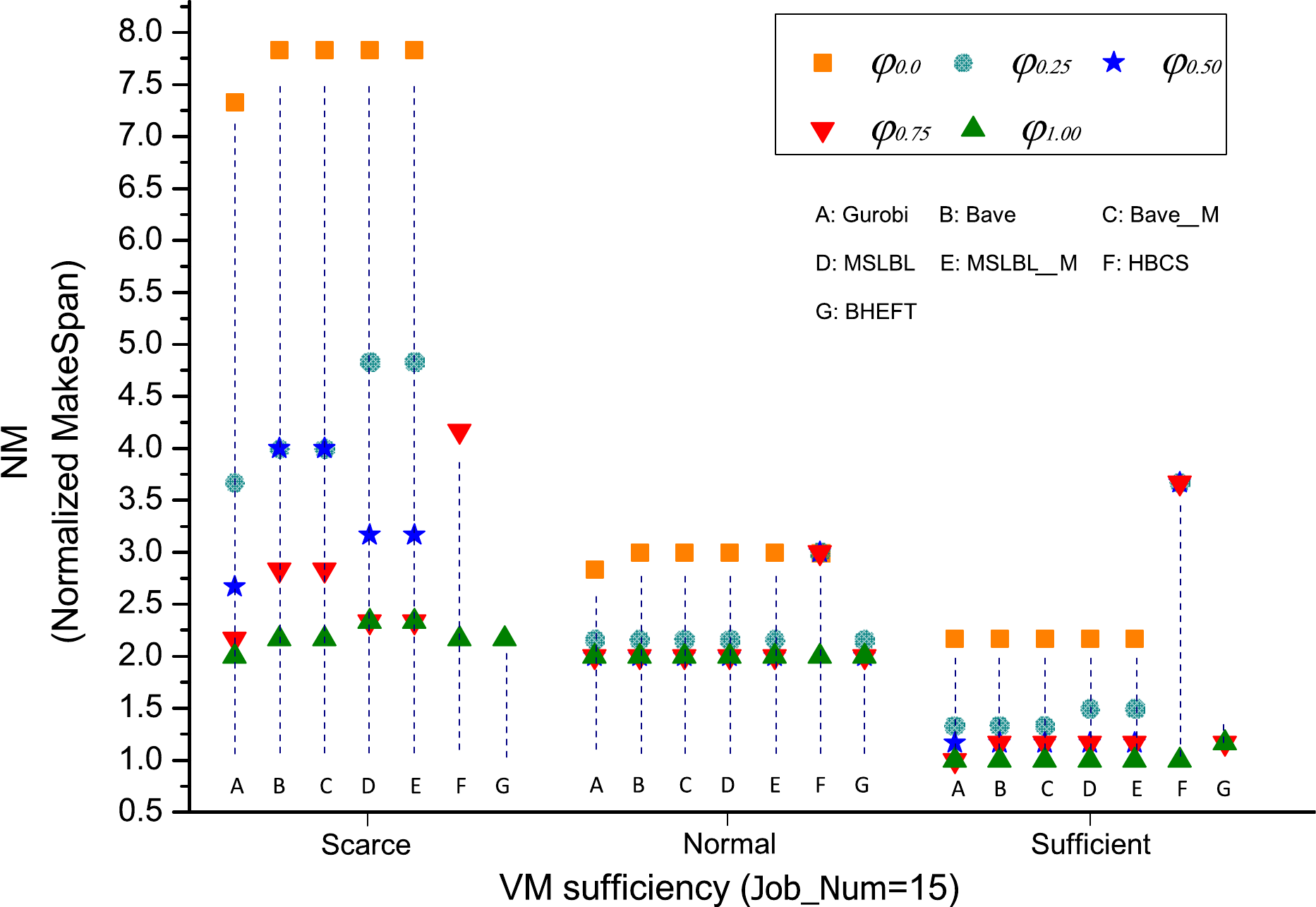} \\ (a) $N = 15$
\end{center}
\end{minipage}
\begin{minipage}{3.5in}
\begin{center}
\includegraphics[width=3.5in]{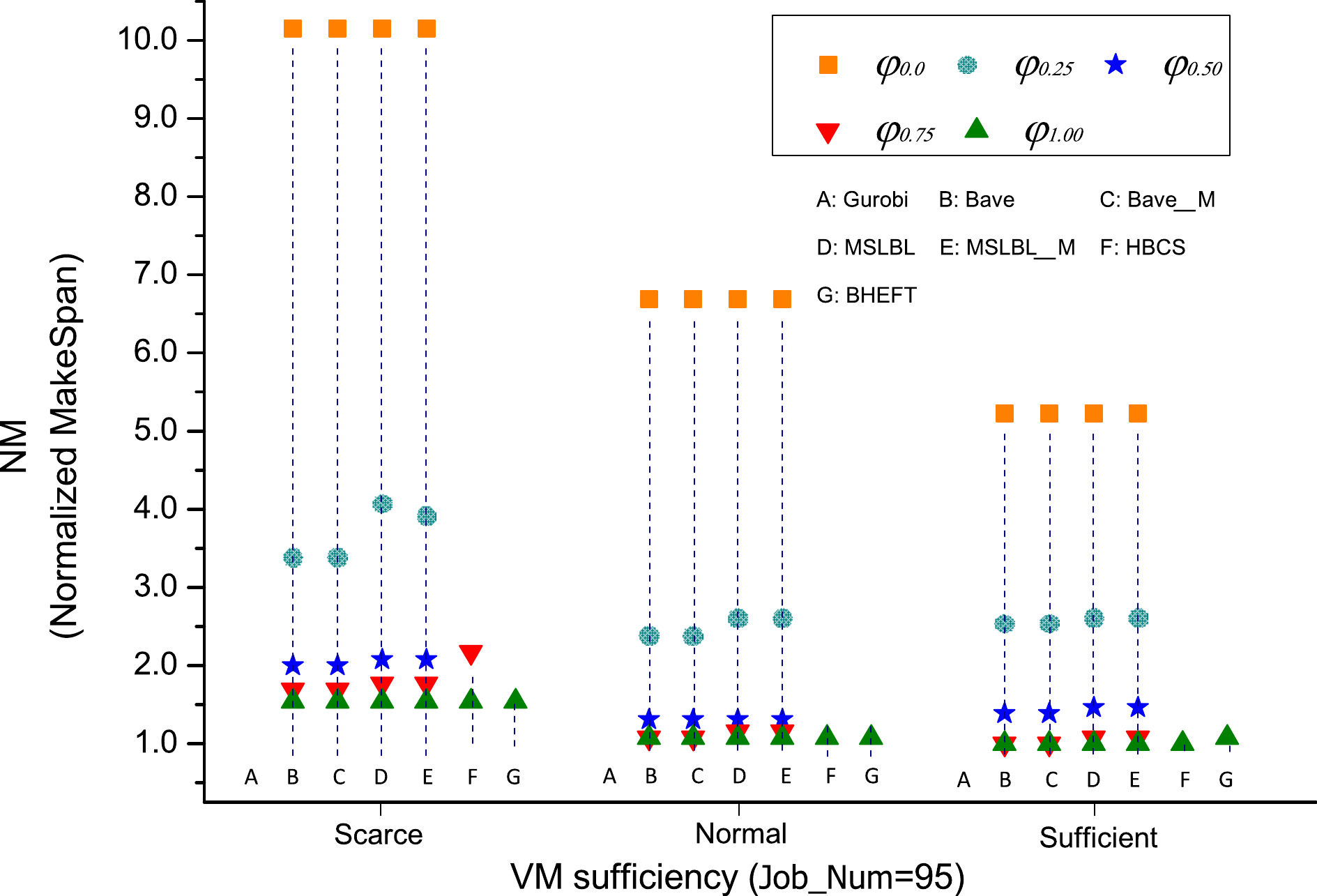} \\ (b) $N = 95$
\end{center}
\end{minipage}
\begin{minipage}{3.5in}
\begin{center}
\includegraphics[width=3.5in]{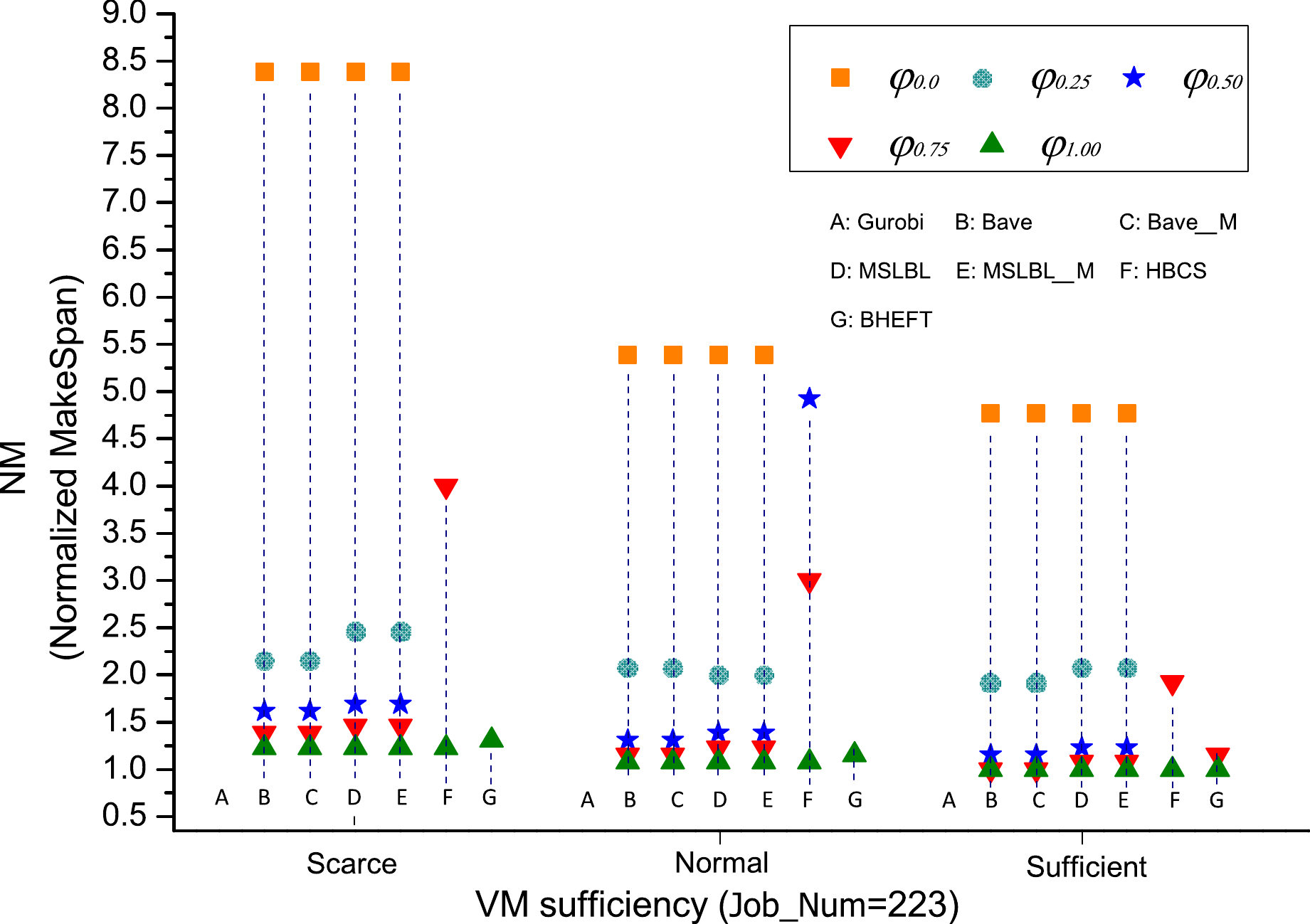} \\ (c) $N = 223$
\end{center}
\end{minipage}
\end{center}
\caption{The normalized makespan with various FFT workflows; $N$ is the number of jobs.}
\end{figure}

\begin{figure}[H]
\ContinuedFloat
\begin{center}
\begin{minipage}{3.5in}
\begin{center}
\includegraphics[width=3.5in]{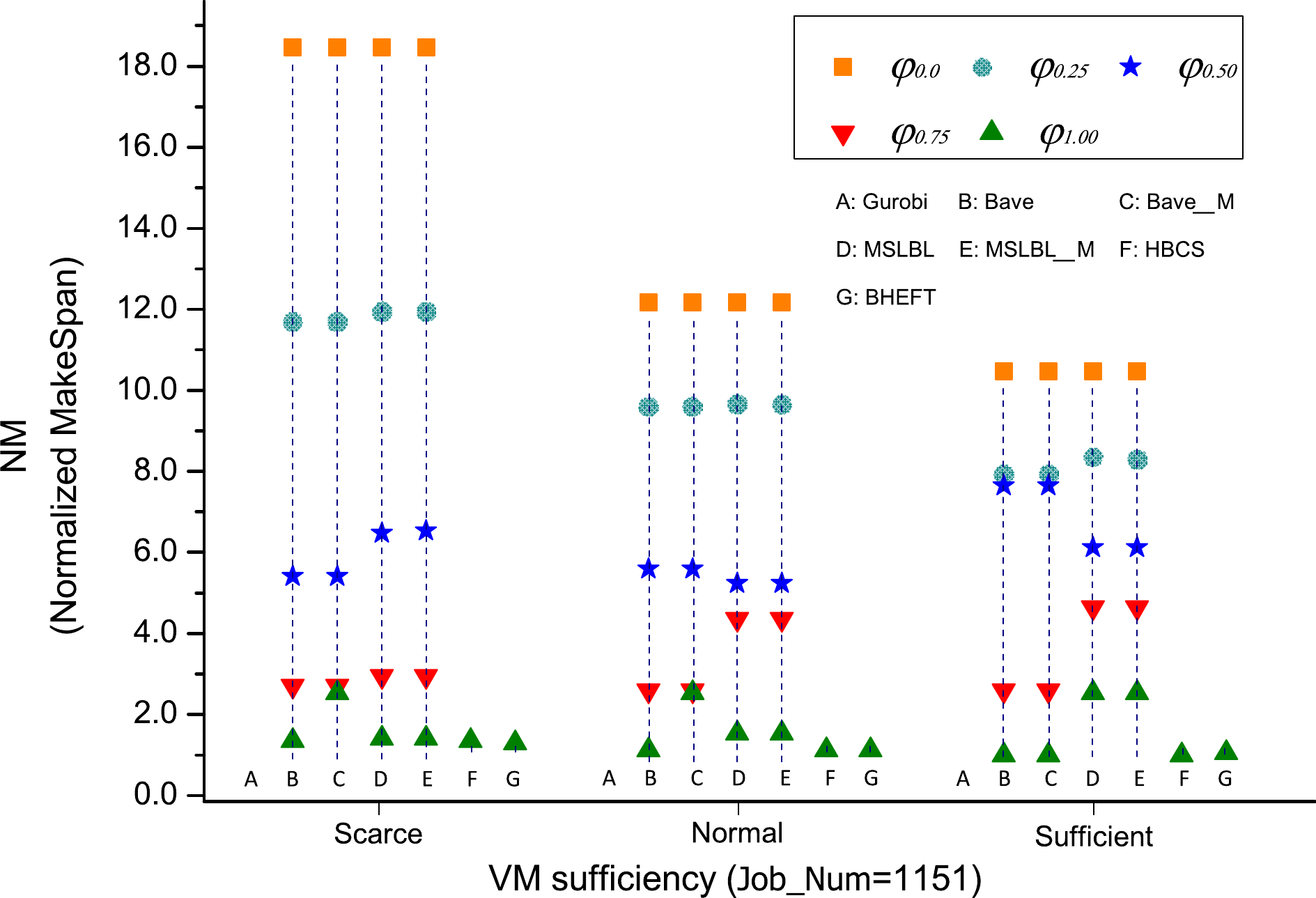} \\ (d) $N = 1151$
\end{center}
\end{minipage}
\begin{minipage}{3.5in}
\begin{center}
\includegraphics[width=3.5in]{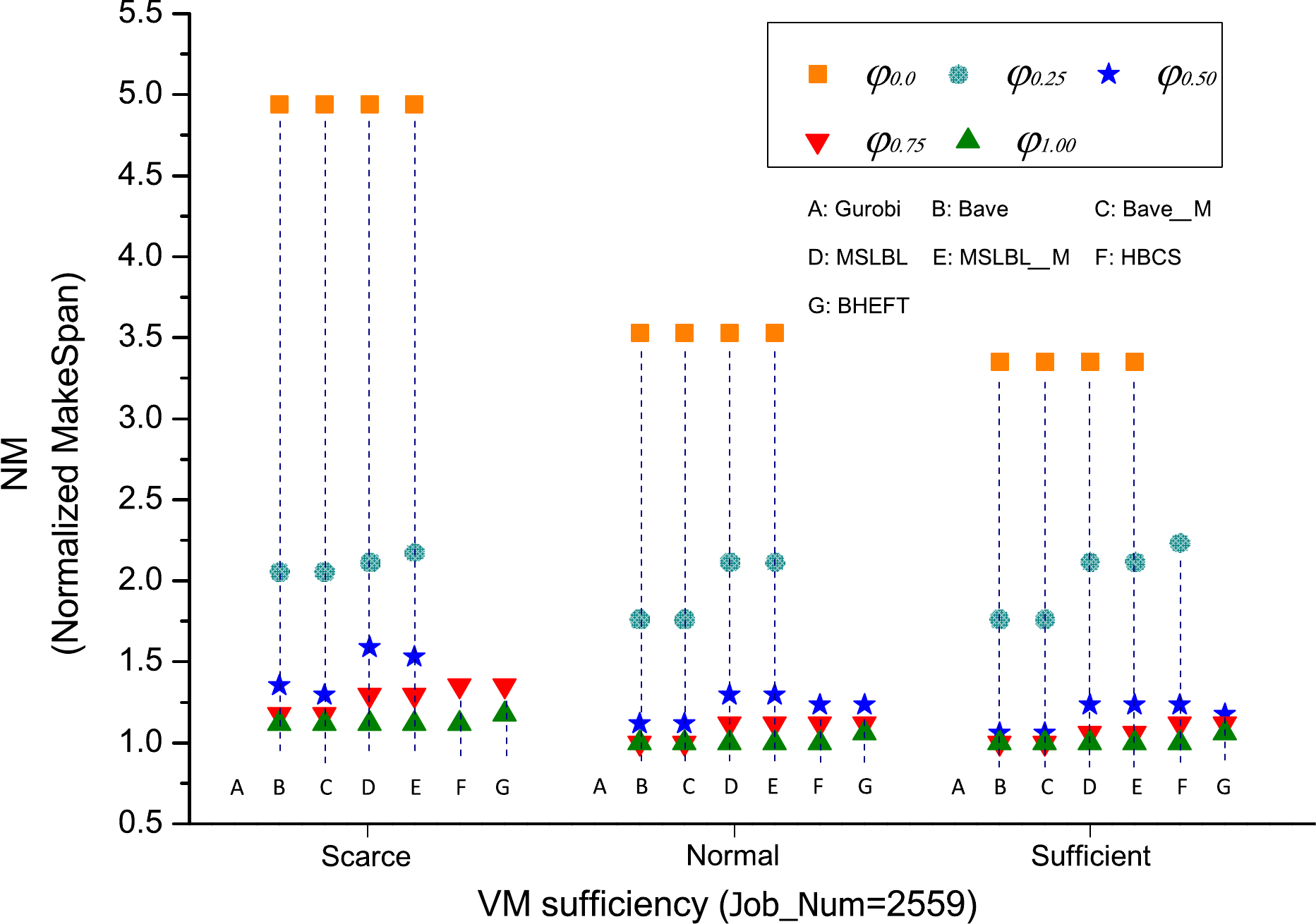} \\ (e) $N = 2559$
\end{center}
\end{minipage}
\end{center}
\caption{(Cont.) The normalized makespan with various FFT workflows; $N$ is the number of jobs.}
\label{fig:fft_vm_nm}
\end{figure}

\subsubsection{Gaussian elimination}
For the Gaussian elimination workflows, the BAVE and BAVE\_M algorithms always achieve the best makespan
for all test cases with budget level at $\varphi=0.0$ or $\varphi=1.0$. For other budget levels, $\varphi = 0.25, 0.5, 0.75$, BAVE always outperforms MSLBL, and BAVE\_M always outperforms MSLBL\_M. The HBCS and
BHEFT algorithms perform poorly in most cases. For most cases, the algorithms with the weighted upward-rank priority generation scheme achieve better makespan. Overall, the proposed BAVE and BAVE\_M work well for Gaussian elimination workflows.

We also observe an unusual test case.
In the test case of $N=1175$, VM {\em Sufficient} and $\varphi = 0.0$, the makespans obtained by BAVE, BAVE\_M, MSLBL, MSLBL\_M are significantly larger than that of the test case where VM sufficiency is {\em Normal}.
This is because that the set of VM instances are generated randomly and can be substantially different for
different VM sufficiency. The increase in VM sufficiency does not necessarily lead to performance improvement.
These kind of rare cases happen occasionally in the other tests.

\begin{figure}[H]
\begin{center}
\begin{minipage}{3.5in}
\begin{center}
\includegraphics[width=3.5in]{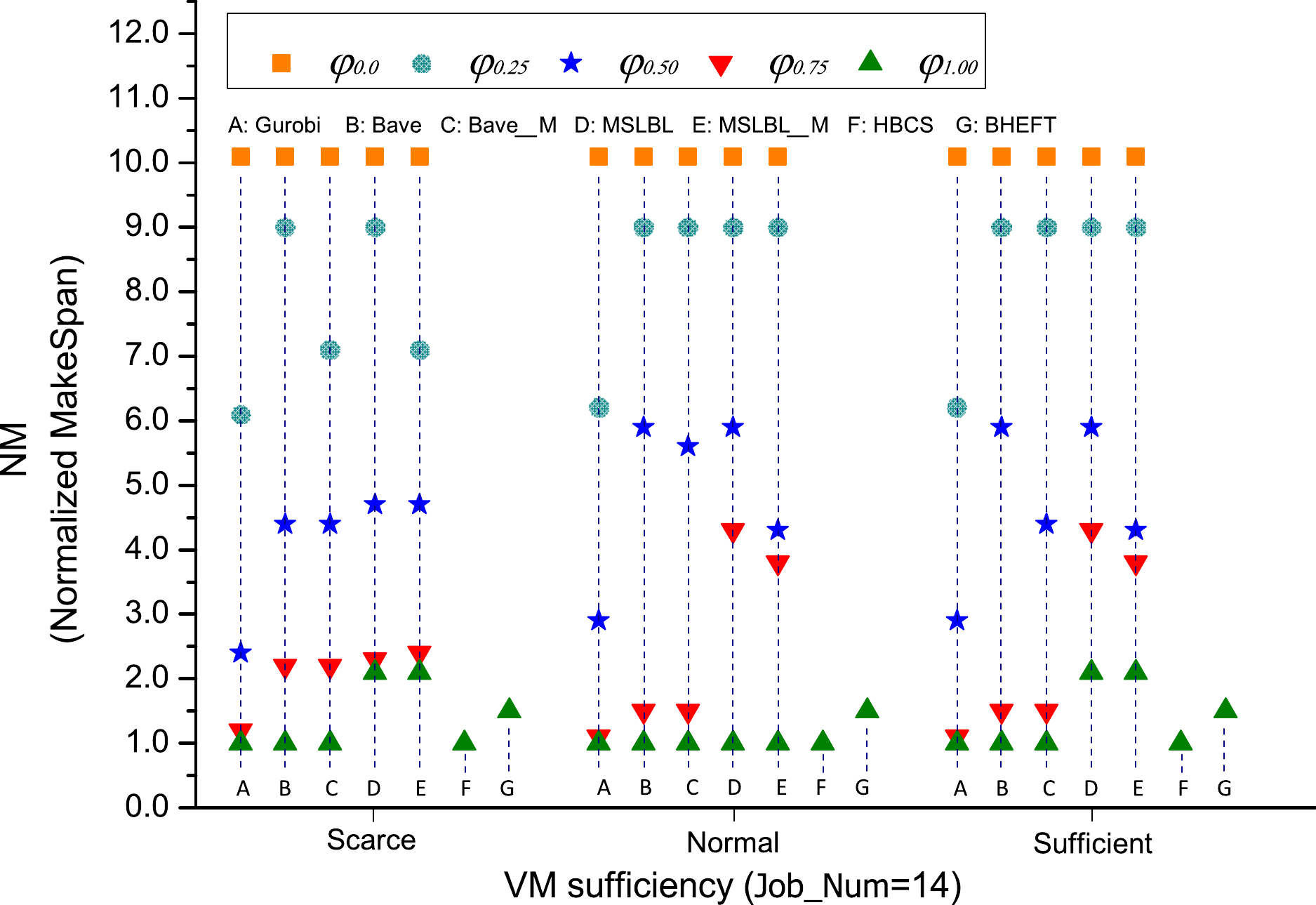} \\ (a) $N = 14$
\end{center}
\end{minipage}
\begin{minipage}{3.5in}
\begin{center}
\includegraphics[width=3.5in]{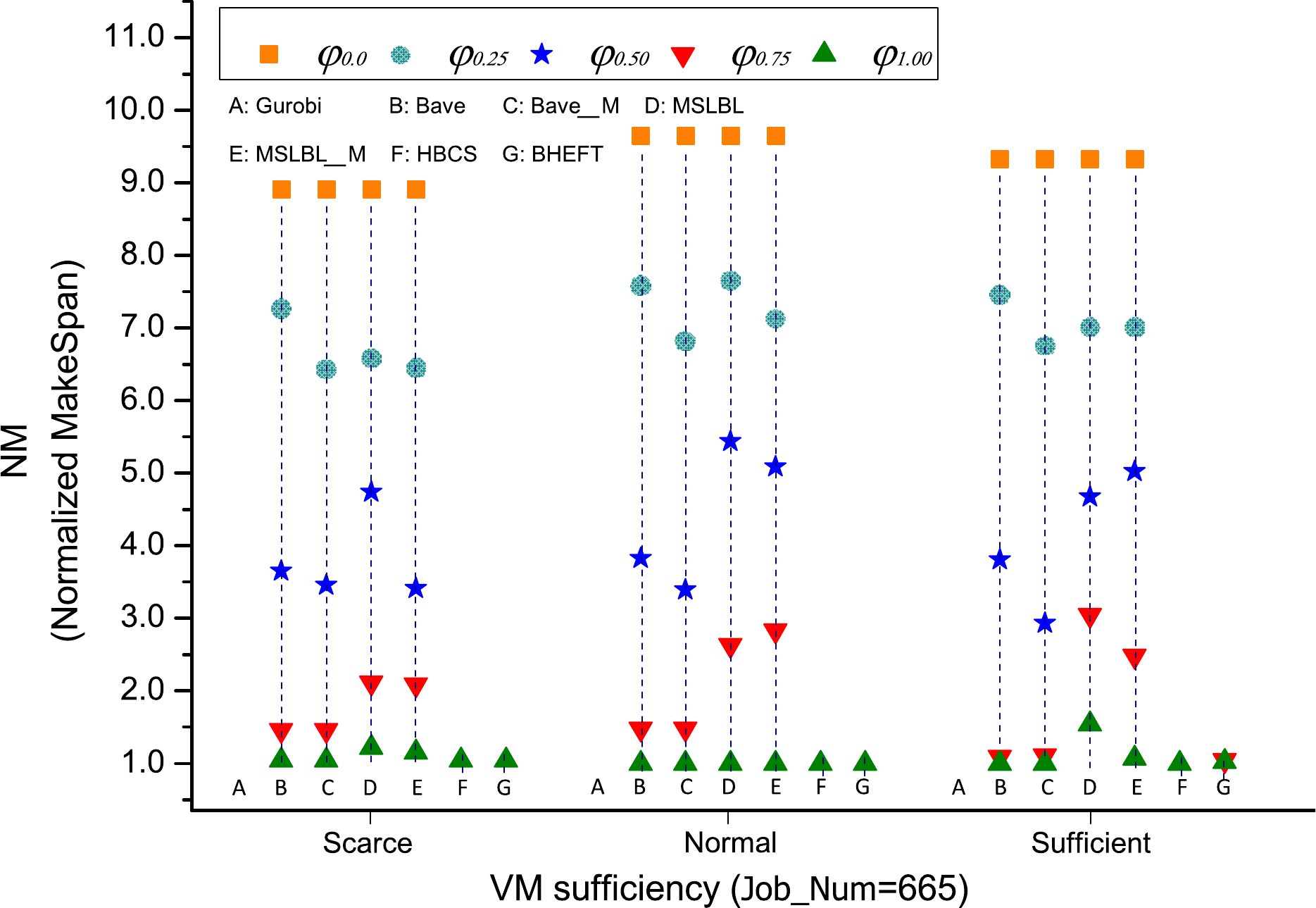} \\ (b) $N = 665$
\end{center}
\end{minipage}
\begin{minipage}{3.5in}
\begin{center}
\includegraphics[width=3.5in]{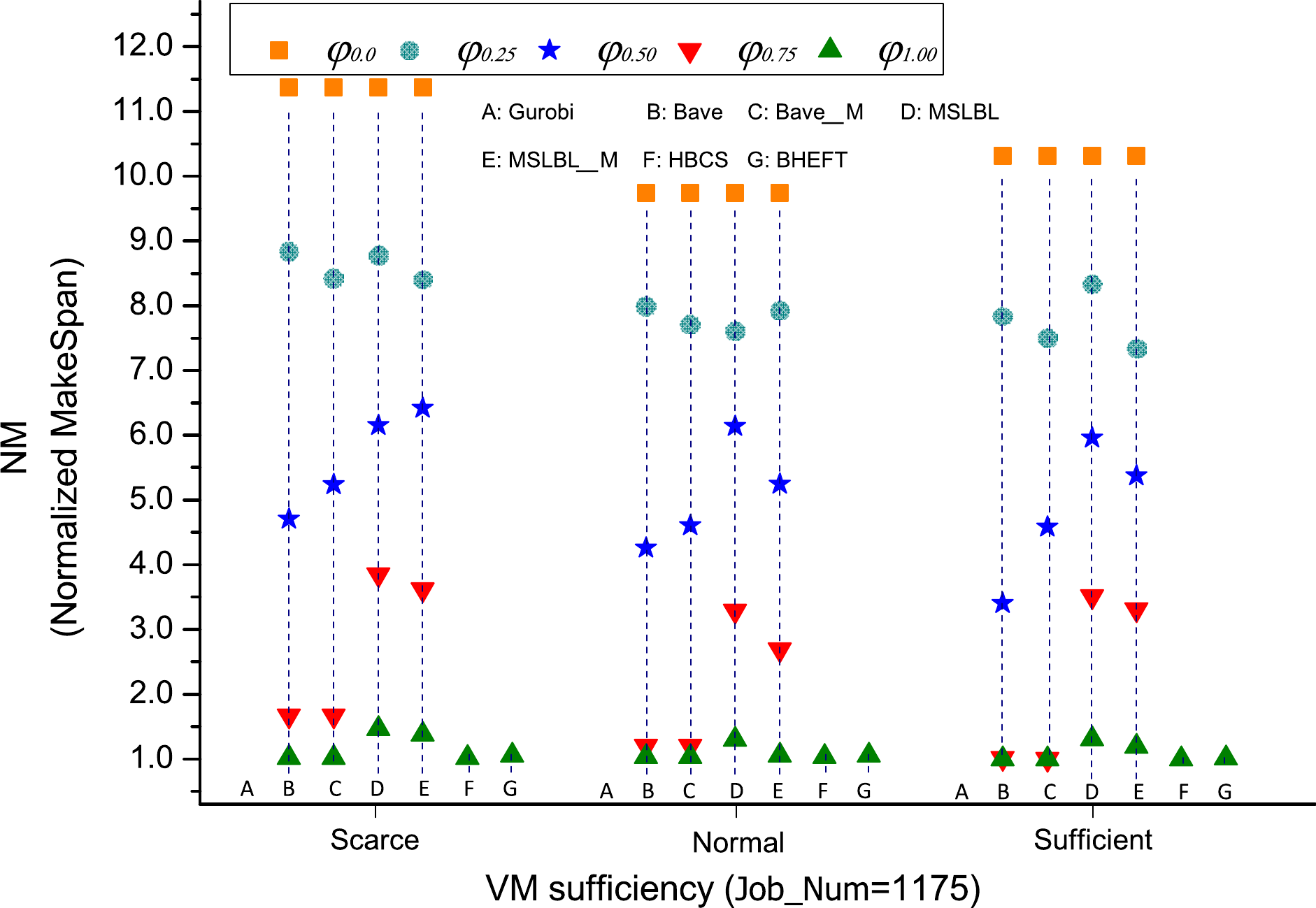} \\ (c) $N = 1175$
\end{center}
\end{minipage}
\end{center}
\caption{The normalized makespan with various Guassian elimination workflows; $N$ is the number of jobs.}
\end{figure}

\begin{figure}[H]
\ContinuedFloat
\begin{center}
\begin{minipage}{3.5in}
\begin{center}
\includegraphics[width=3.5in]{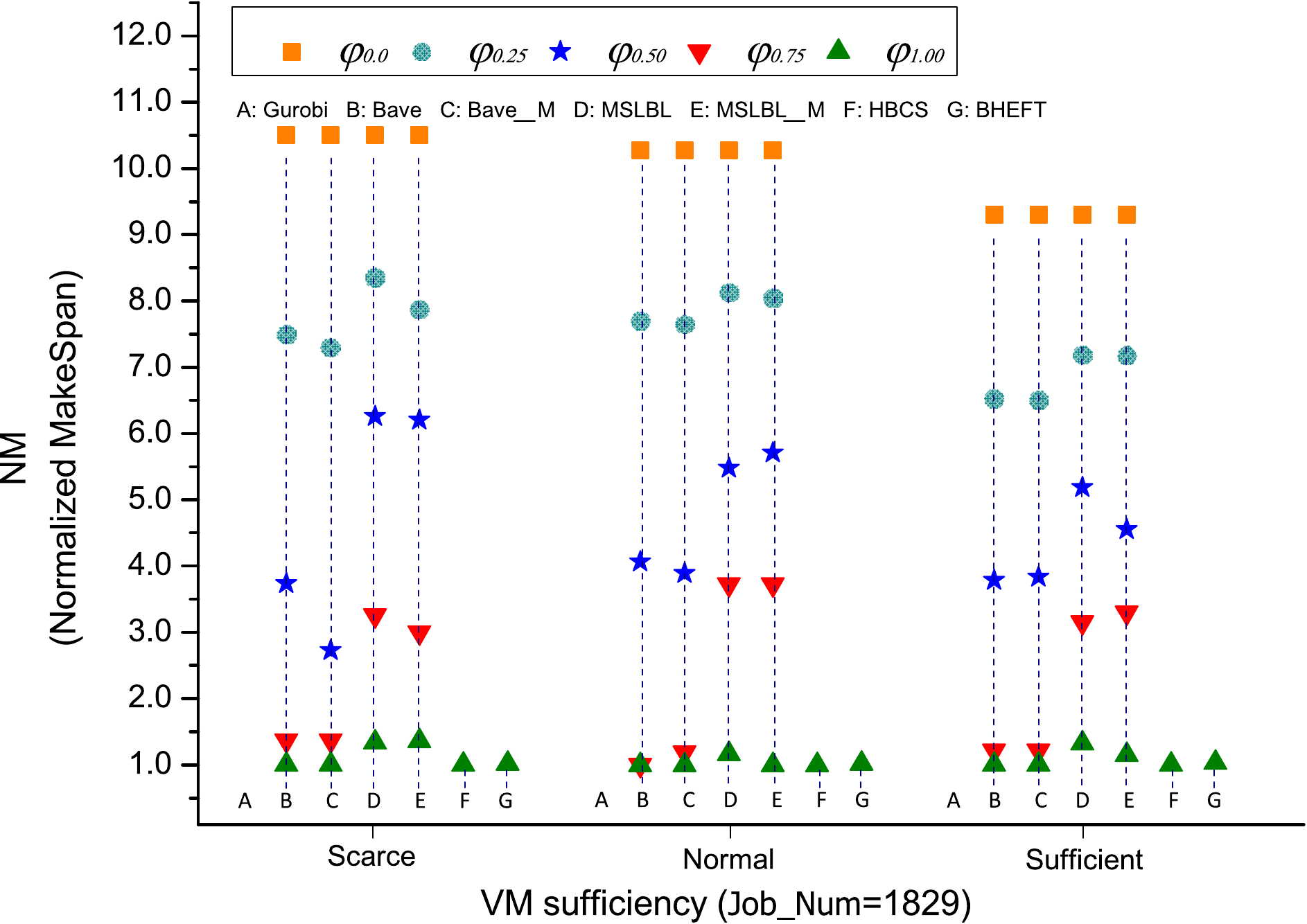} \\ (d) $N = 1829$
\end{center}
\end{minipage}
\end{center}
\caption{(Cont.) The normalized makespan with various Guassian elimination workflows; $N$ is the number of jobs.}
\label{fig:guassian_vm_nm}
\end{figure}

\begin{figure}[H]
\begin{center}
\begin{minipage}{3.5in}
\begin{center}
\includegraphics[width=3.5in]{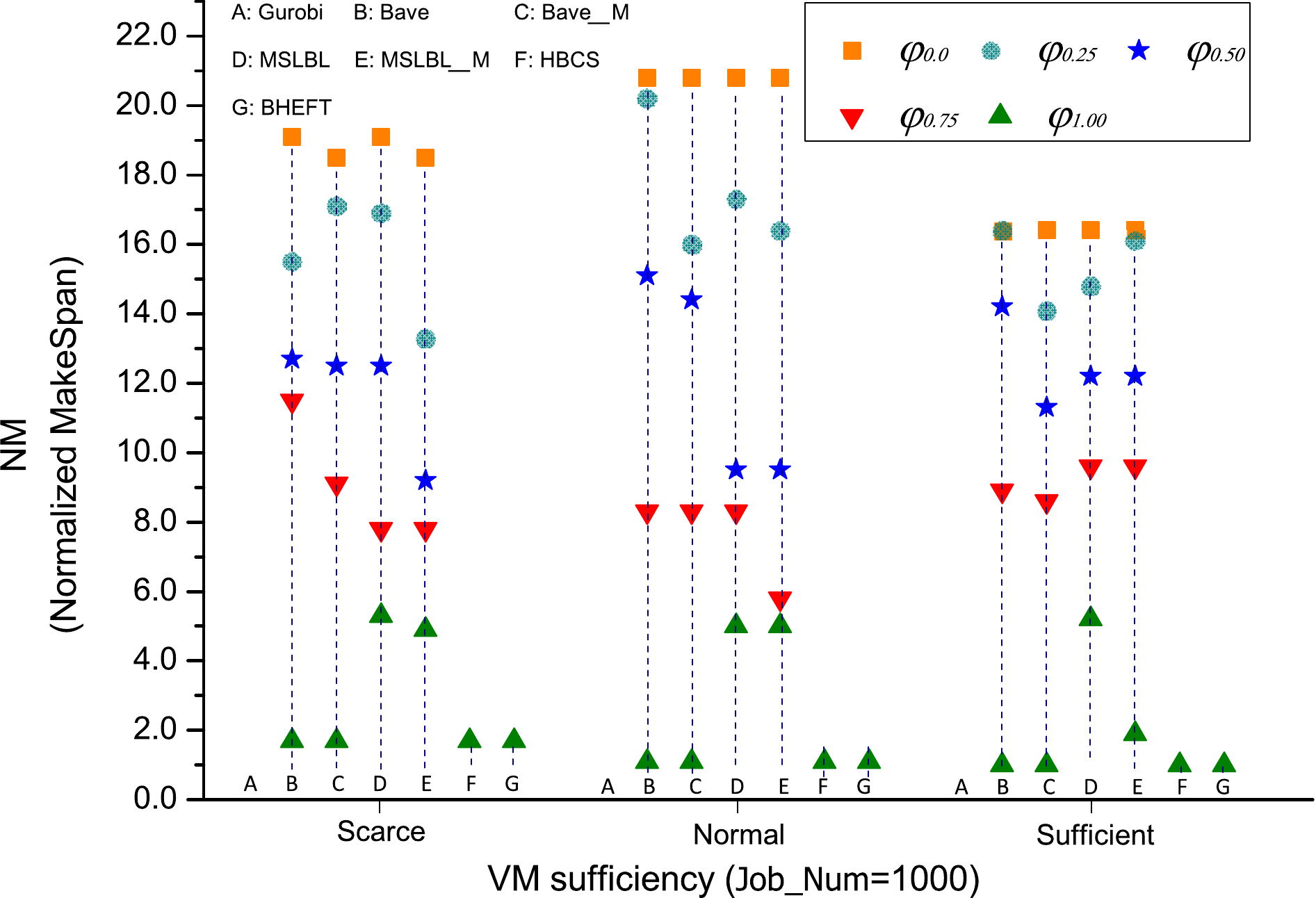} \\ (a) CyberShake with $N = 1000$
\end{center}
\end{minipage}
\begin{minipage}{3.5in}
\begin{center}
\includegraphics[width=3.5in]{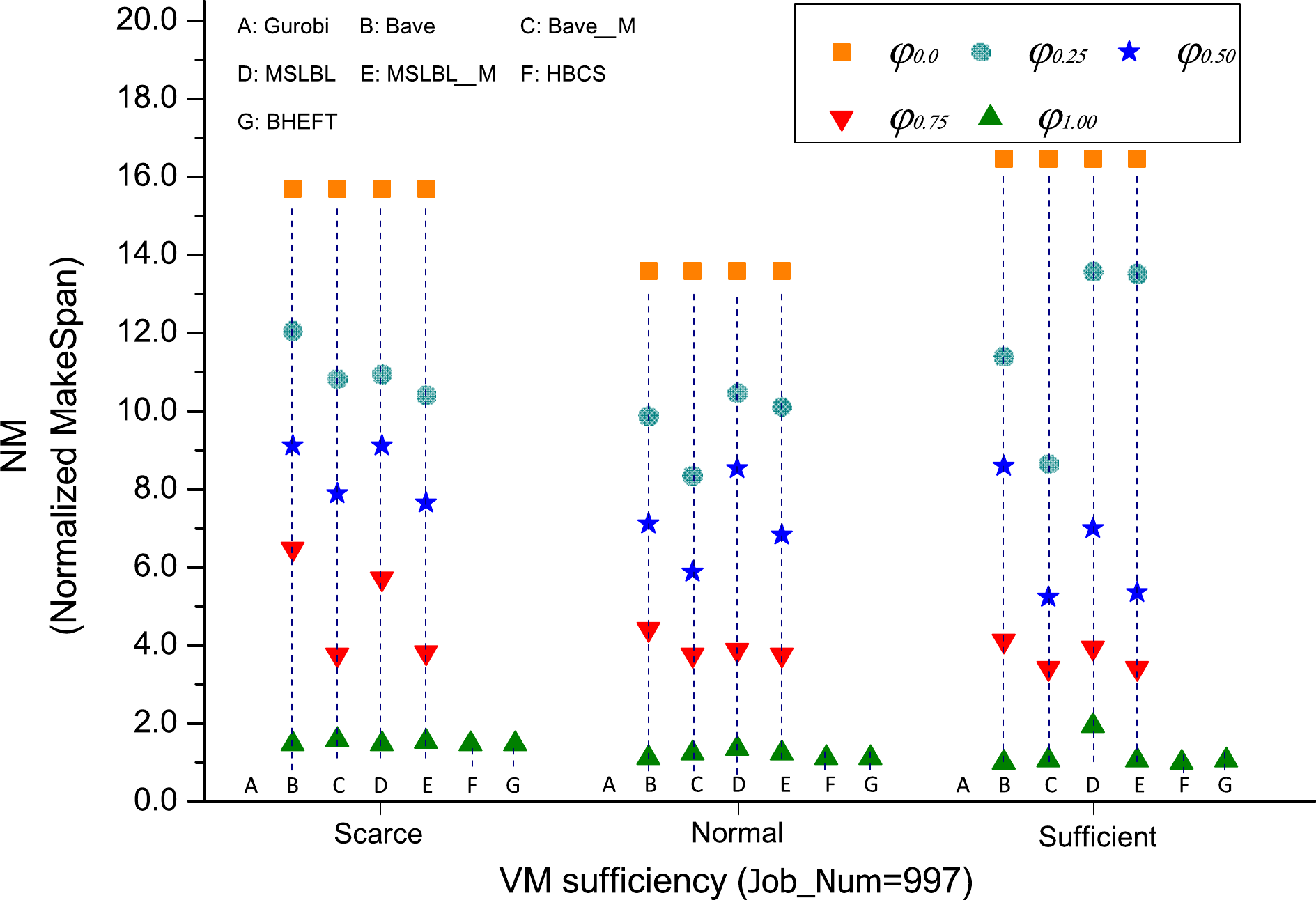} \\ (b) Epigenomics with $N = 997$
\end{center}
\end{minipage}
\end{center}
\caption{The normalized makespan with other scientific workflows; $N$ is the number of jobs.}
\end{figure}

\begin{figure}[H]
\ContinuedFloat
\begin{center}
\begin{minipage}{3.5in}
\begin{center}
\includegraphics[width=3.5in]{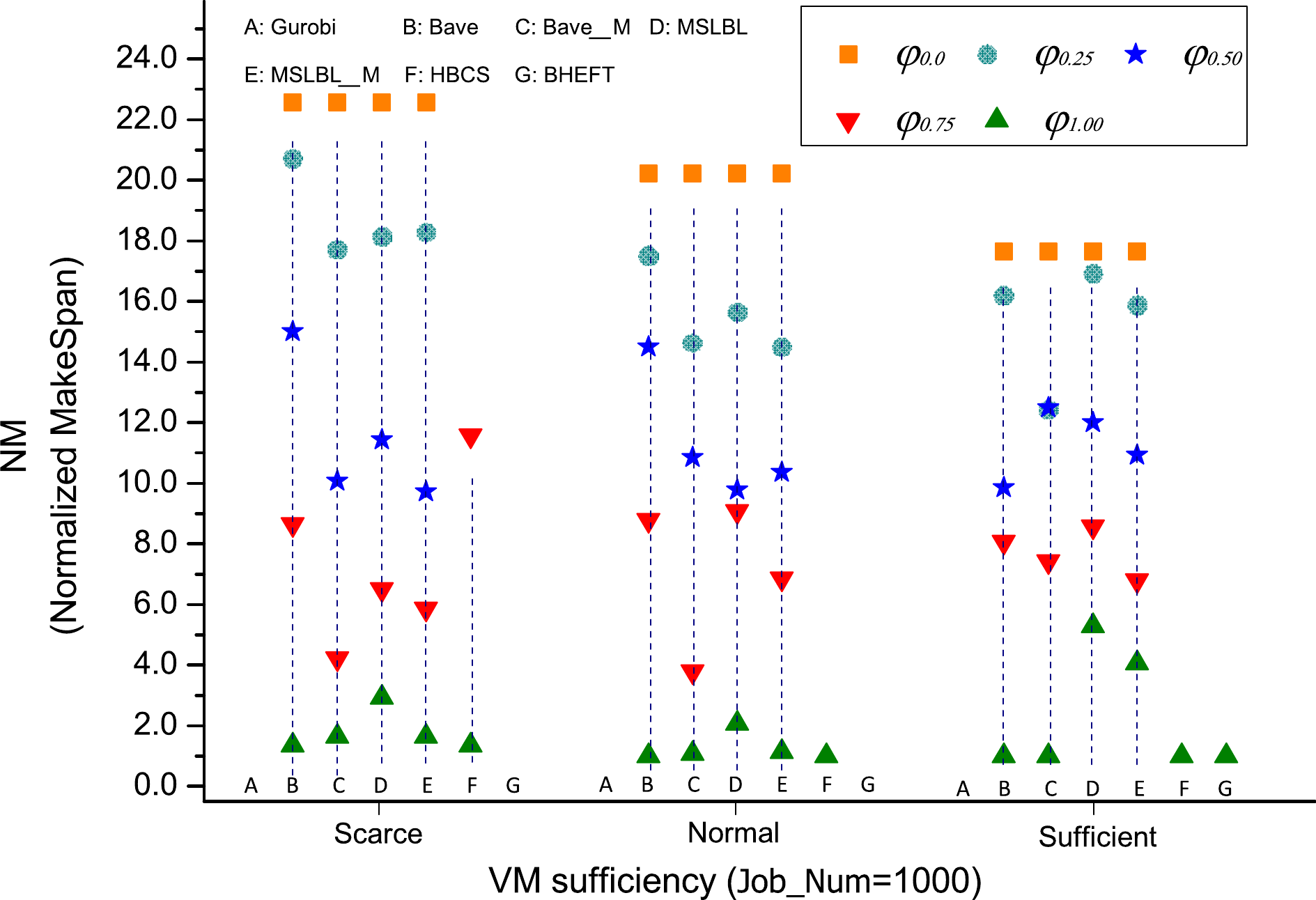} \\ (c) Inspiral with $N = 1000$
\end{center}
\end{minipage}
\begin{minipage}{3.5in}
\begin{center}
\includegraphics[width=3.5in]{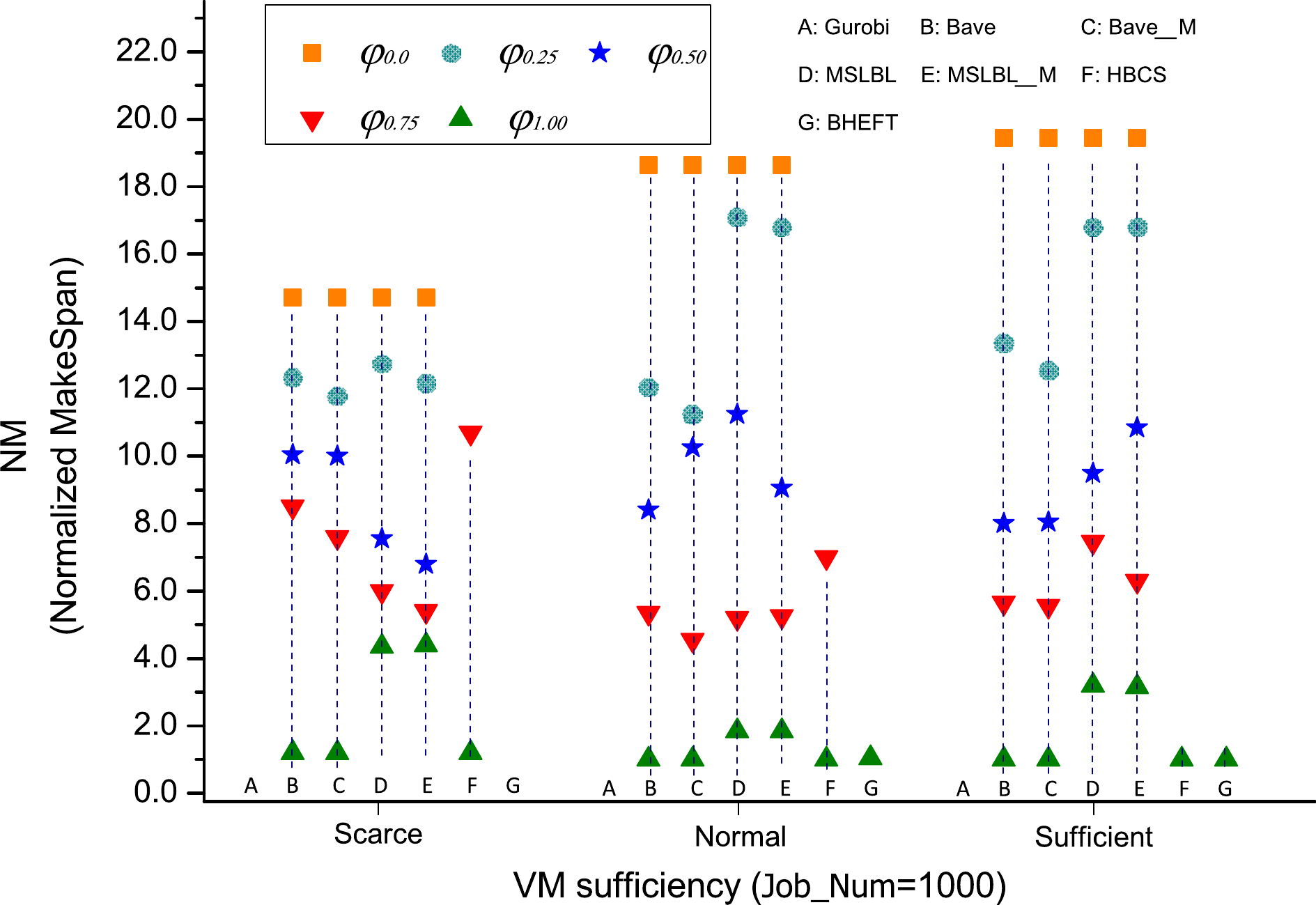} \\ (d) Montage with $N = 1000$
\end{center}
\end{minipage}
\begin{minipage}{3.5in}
\begin{center}
\includegraphics[width=3.5in]{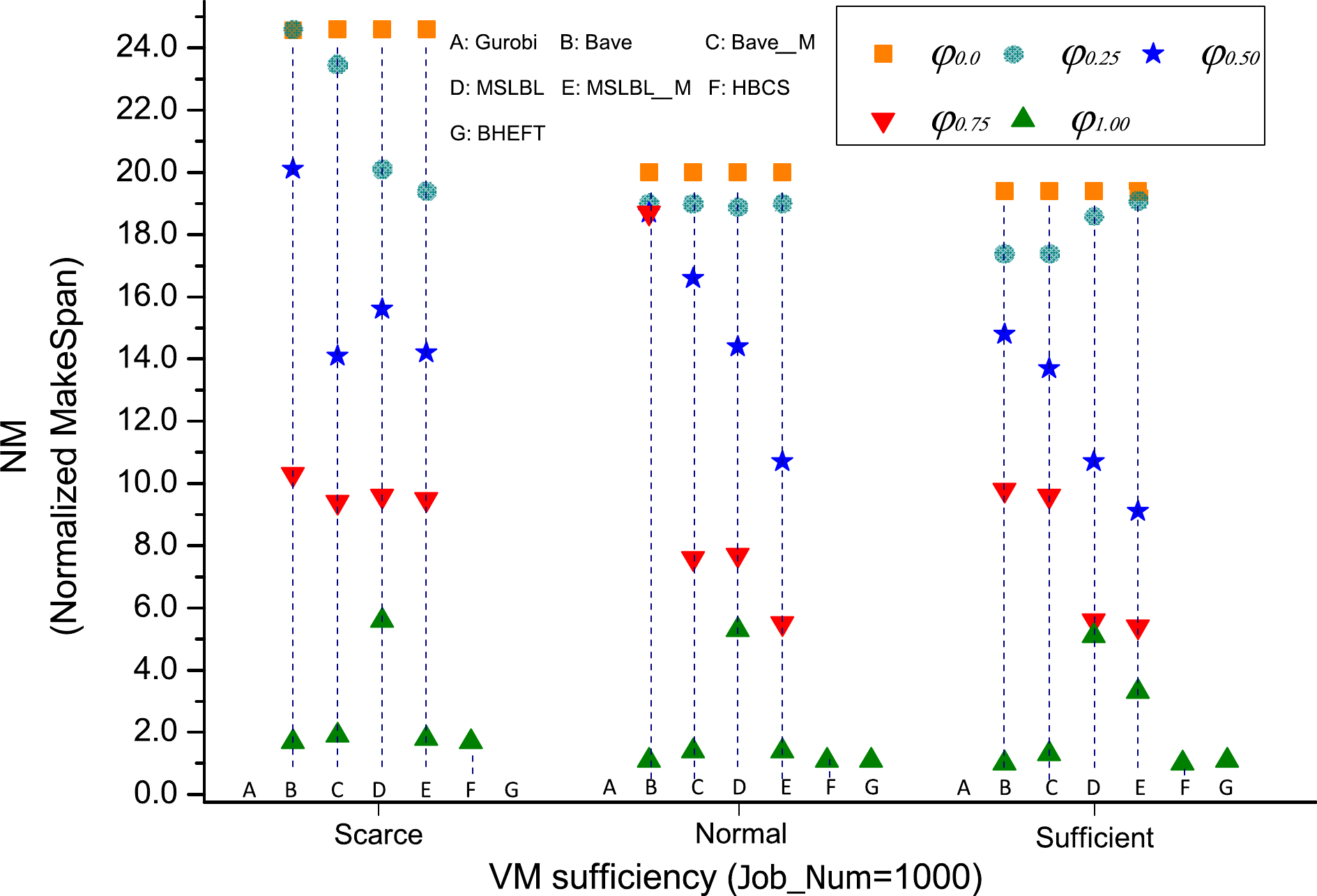} \\ (e) Sipht with $N = 1000$
\end{center}
\end{minipage}
\end{center}
\caption{(Cont.) The normalized makespan with other scientific workflows; $N$ is the number of jobs.}
\label{fig:scientific_vm_nm}
\end{figure}

\subsubsection{Other scientific workflows}
We show the evaluation results of other scientific workflows in Fig. \ref{fig:scientific_vm_nm}.
For the CyberShake workflow with $N = 1000$ jobs, the BAVE algorithm performs no better than MSLBL. When it is applied with the weighted
priority scheme, the makespans are reduced and BAVE\_M performs the best in average. For the Sipht workflow, though the BAVE and BAVE\_M algorithms perform no better than MSLBL and MSLBL\_M, the weighted priority scheme produces shorter makespans than the plain one in most cases. For other kinds of workflows, the BAVE and BAVE\_M algorithms perform the best in most test cases. The performance of HBCS and BHEFT is poor when the budget is not sufficient.

\begin{figure}[H]
\begin{center}
\begin{minipage}{3.5in}
\begin{center}
\includegraphics[width=3.5in]{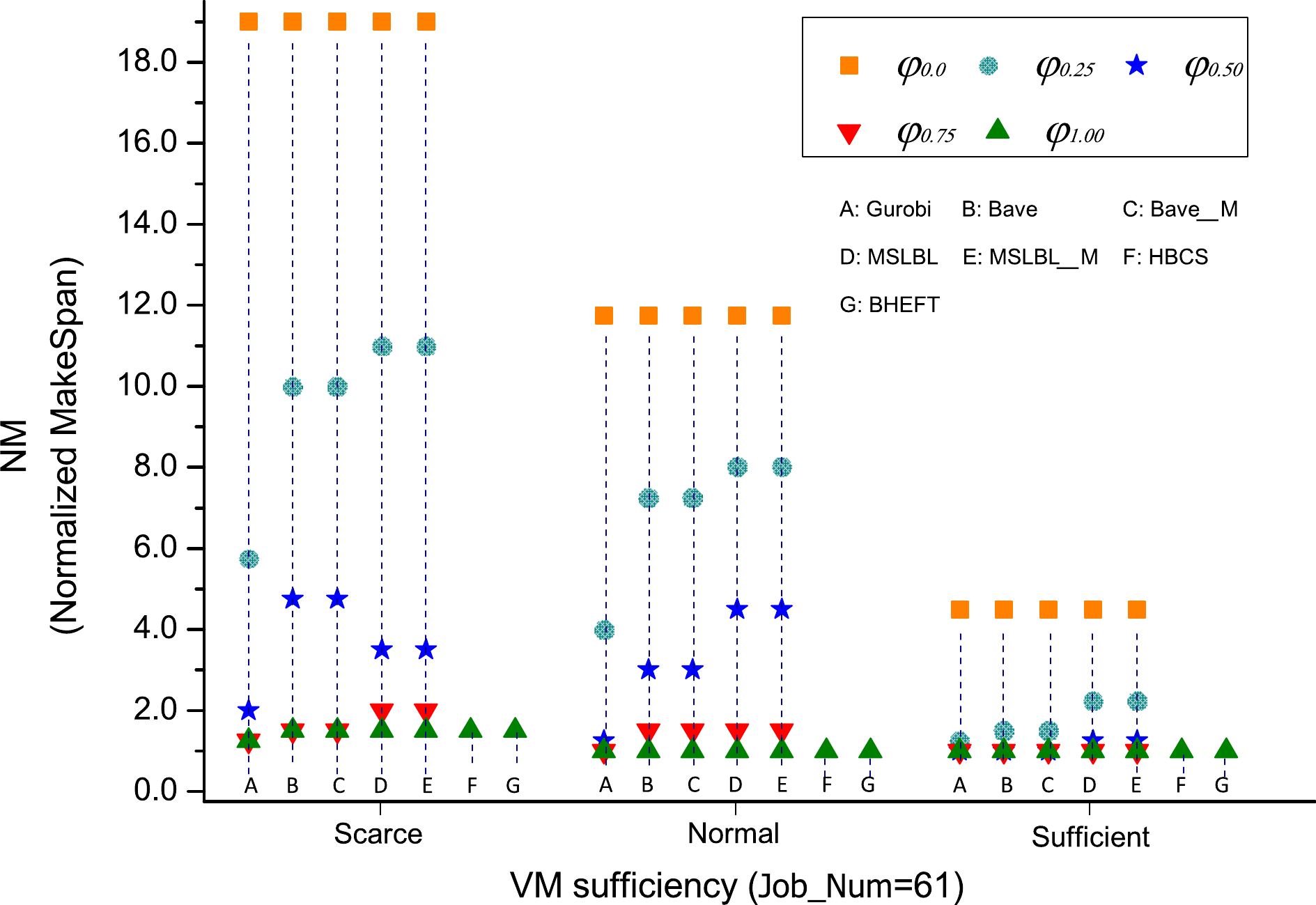} \\ (a) $N = 61$
\end{center}
\end{minipage}
\begin{minipage}{3.5in}
\begin{center}
\includegraphics[width=3.5in]{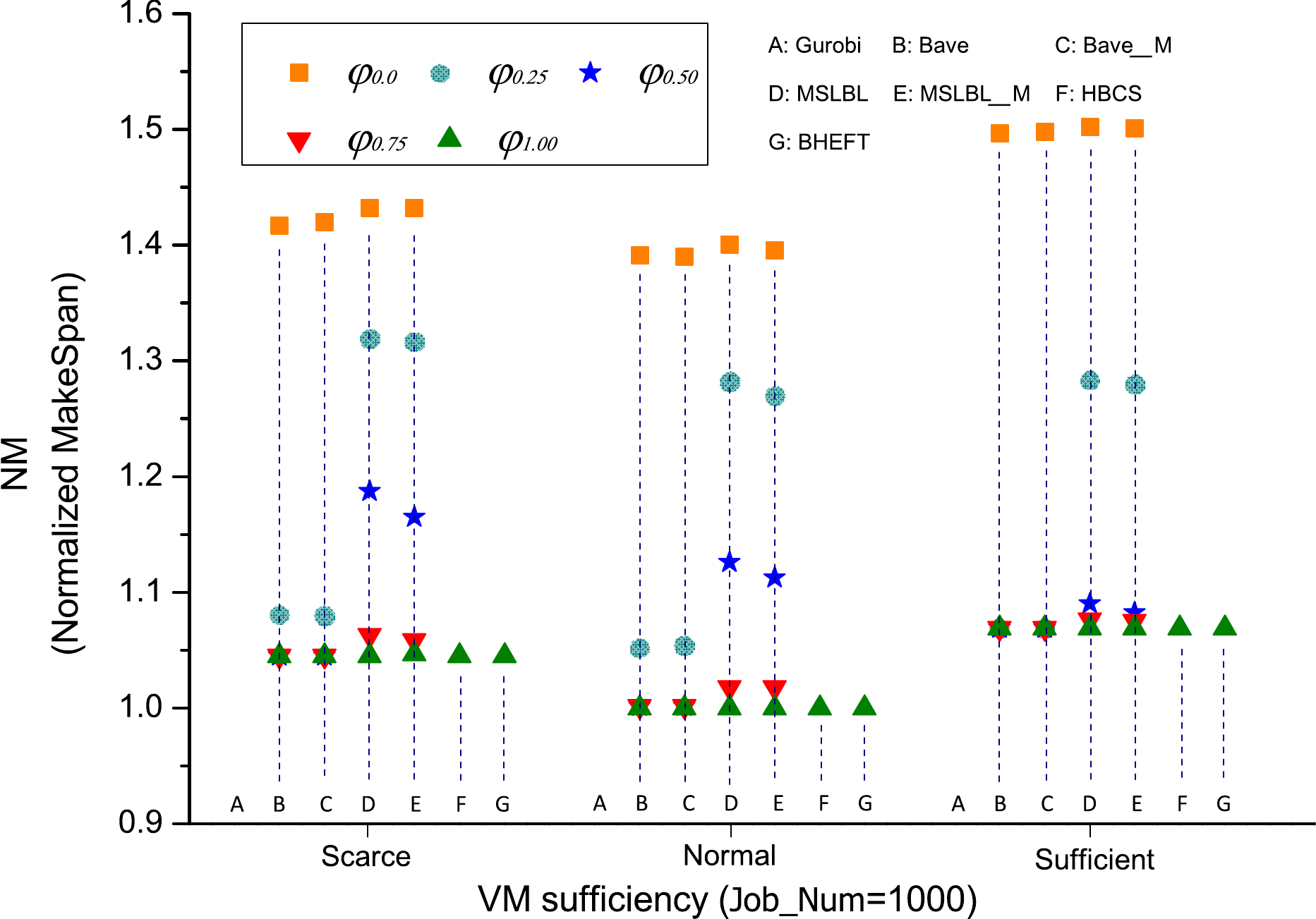} \\ (b) $N = 1000$
\end{center}
\end{minipage}
\begin{minipage}{3.5in}
\begin{center}
\includegraphics[width=3.5in]{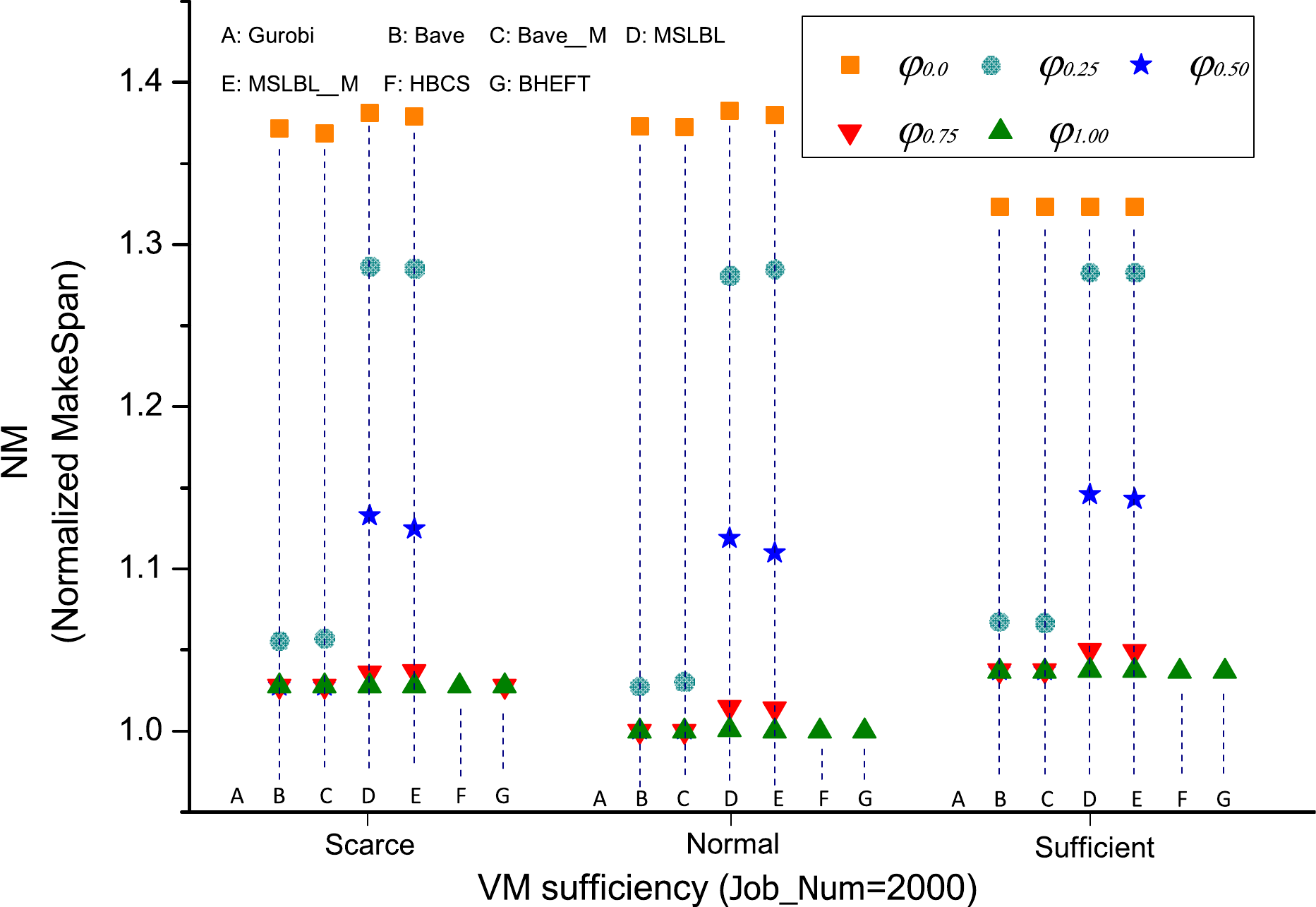} \\ (c) $N = 2000$
\end{center}
\end{minipage}
\end{center}
\caption{The normalized makespan with various randomly generated workflows; $N$ is the number of jobs.}
\label{fig:random_vm_nm}
\end{figure}

\subsubsection{Randomly generated workflows}
We tested various randomly generated workflows and the main results are shown in Fig. \ref{fig:random_vm_nm}. Both the BAVE and BAVE\_M algorithms perform the best compared with other algorithms. There is no significant
performance difference between the plain and weighted priority schemes.

\begin{figure}[H]
\begin{center}
\begin{minipage}{3.5in}
\begin{center}
\includegraphics[width=3.5in]{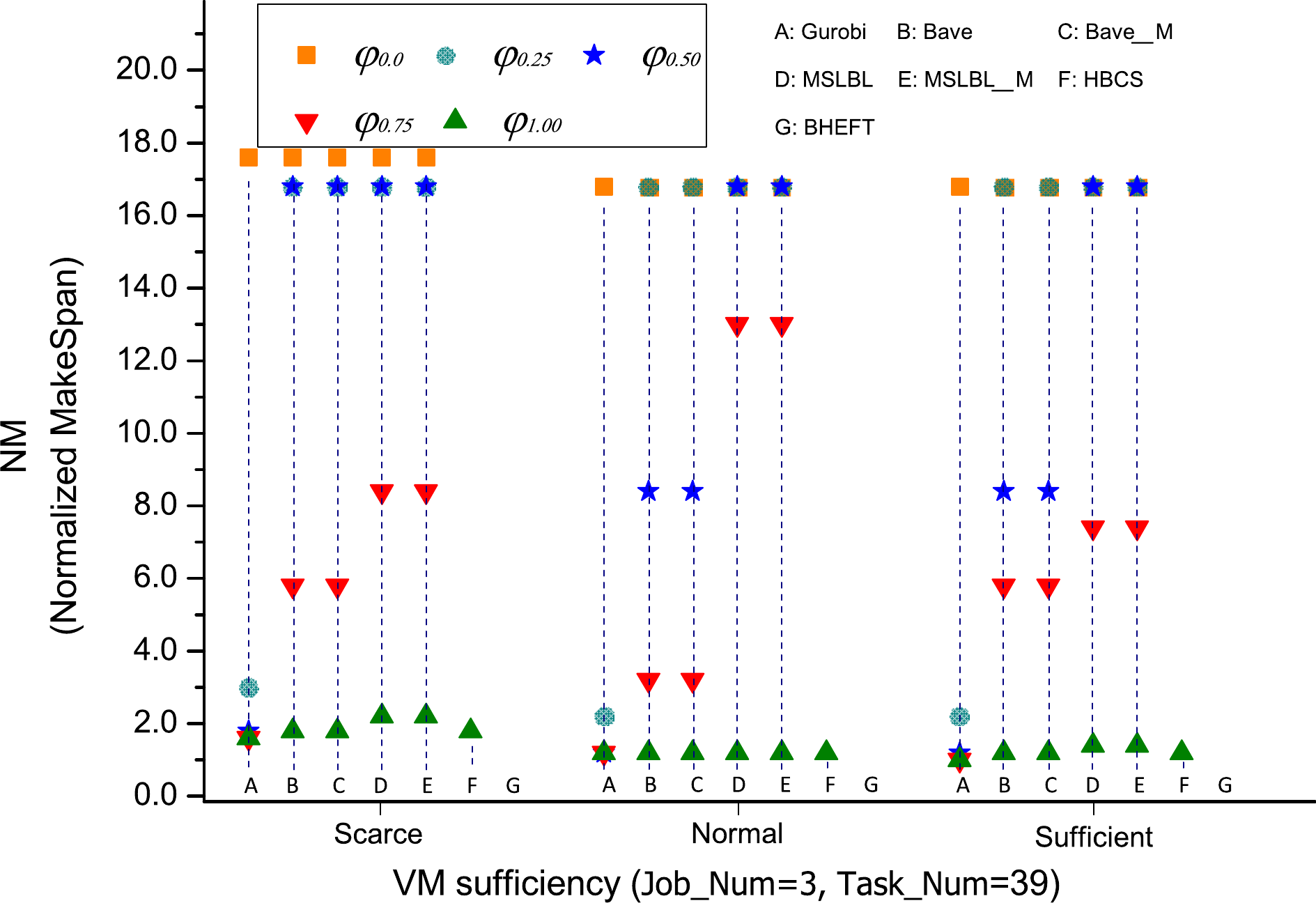} \\ (a) $N = 39$
\end{center}
\end{minipage}
\begin{minipage}{3.5in}
\begin{center}
\includegraphics[width=3.5in]{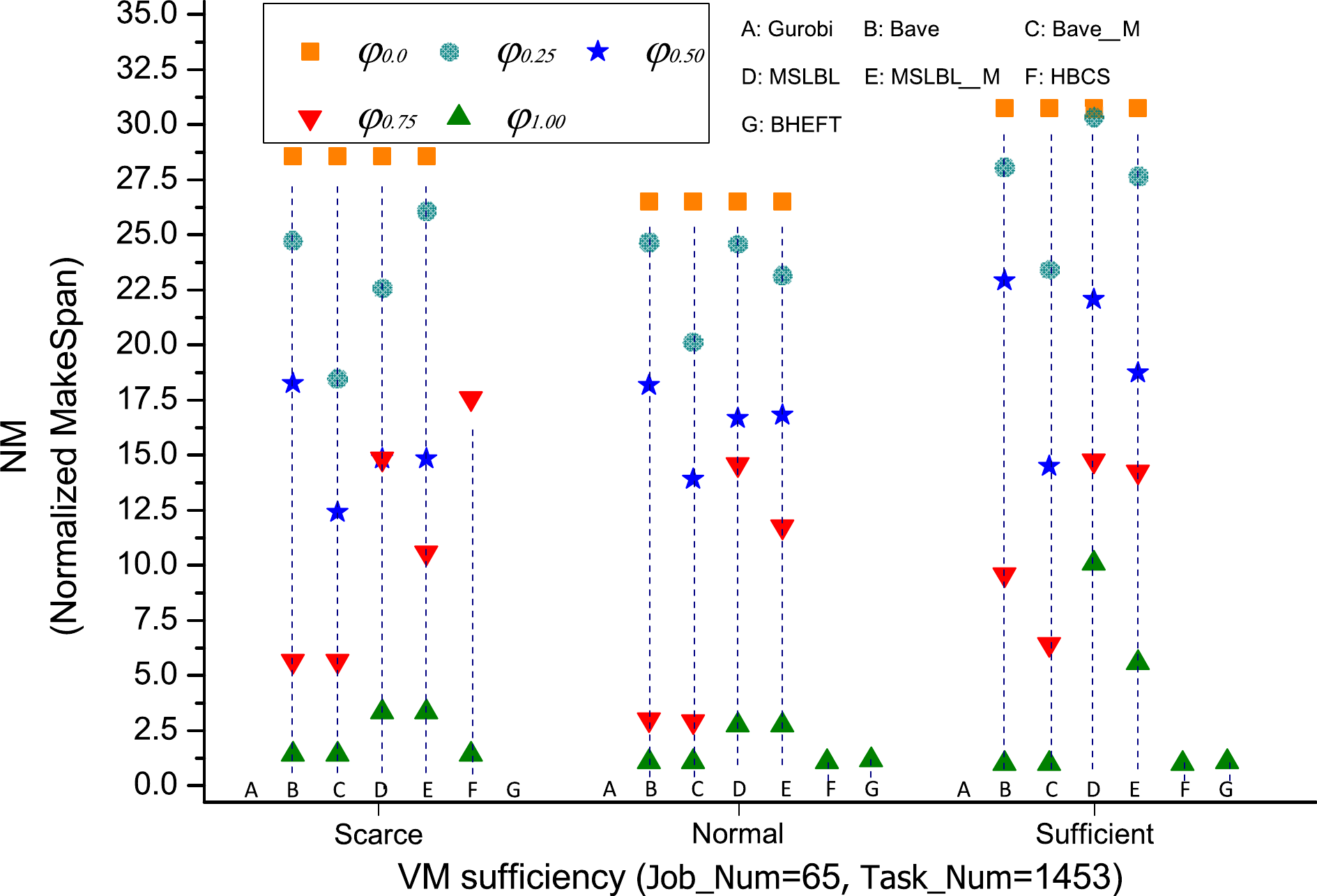} \\ (b) $N = 1453$
\end{center}
\end{minipage}
\begin{minipage}{3.5in}
\begin{center}
\includegraphics[width=3.5in]{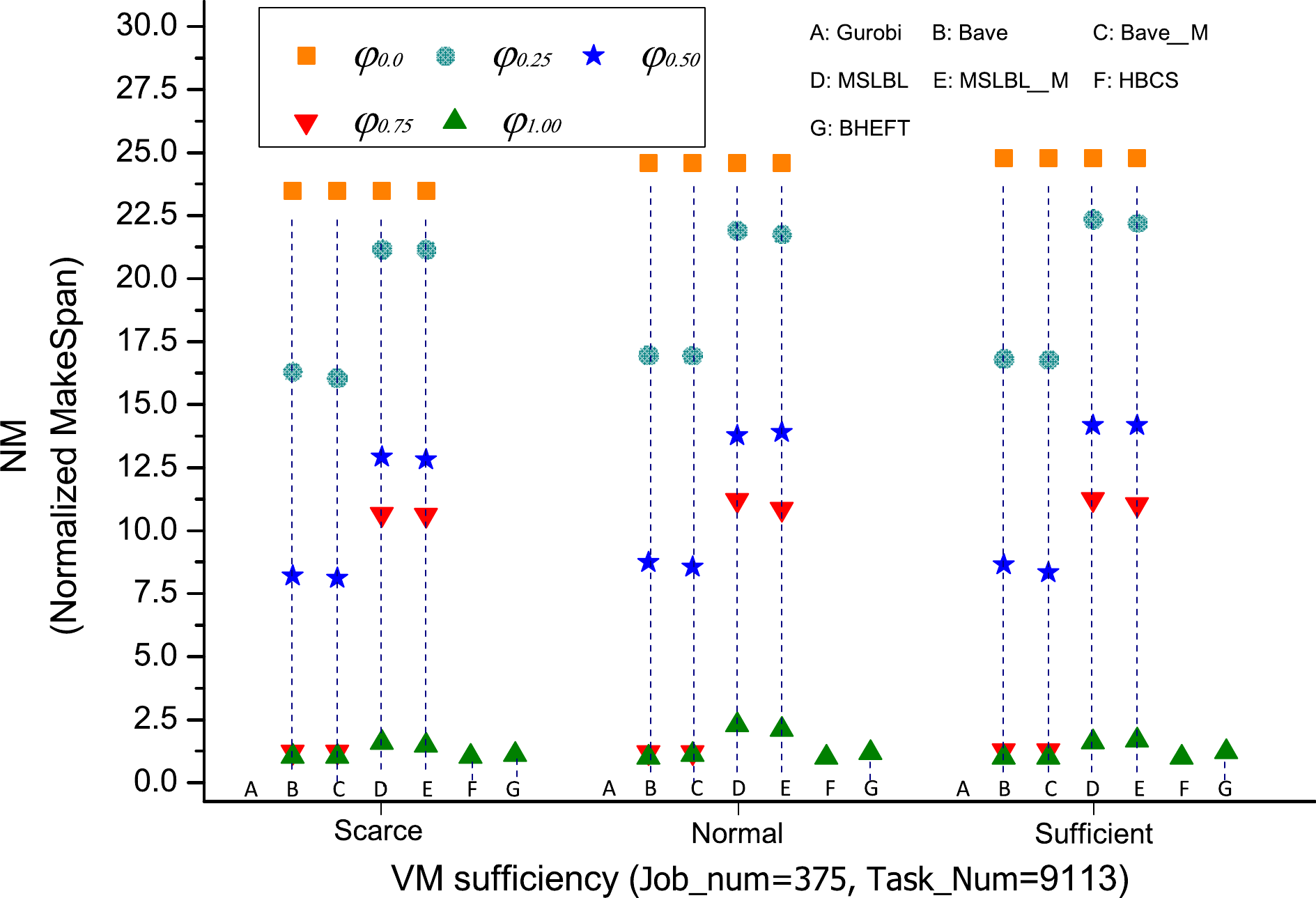} \\ (c) $N = 9113$
\end{center}
\end{minipage}
\end{center}
\caption{The normalized makespan with various workflows from an Internet streaming service company; $N$ is the number of tasks.}
\label{fig:Internet_vm_nm}
\end{figure}

\subsubsection{Workflows from an Internet streaming service company}
Finally, we tested workflows obtained from an Internet streaming service company.
The workflows obtained from the in-production cluster contain multiple jobs,
and each job may contain multiple parallel tasks. Therefore, we evaluate the scale of each workflow by
the number of tasks it carries.
The workflow with $N = 39$ tasks is the
largest test case that Gurobi can solve. The workflow with $N = 1453$ is a medium-sized one that has the highest
occurrence rate among all medium-sized workflows. The workflow with $N = 9113$ is the largest workflow we obtained.
The results in Fig. \ref{fig:Internet_vm_nm} show that BAVE and BAVE\_M outperform other algorithms, and the weighted priority scheme
achieves better performance than the plain one.

\begin{figure}[H]
\begin{center}
\begin{minipage}{3.5in}
\begin{center}
\includegraphics[width=3.5in]{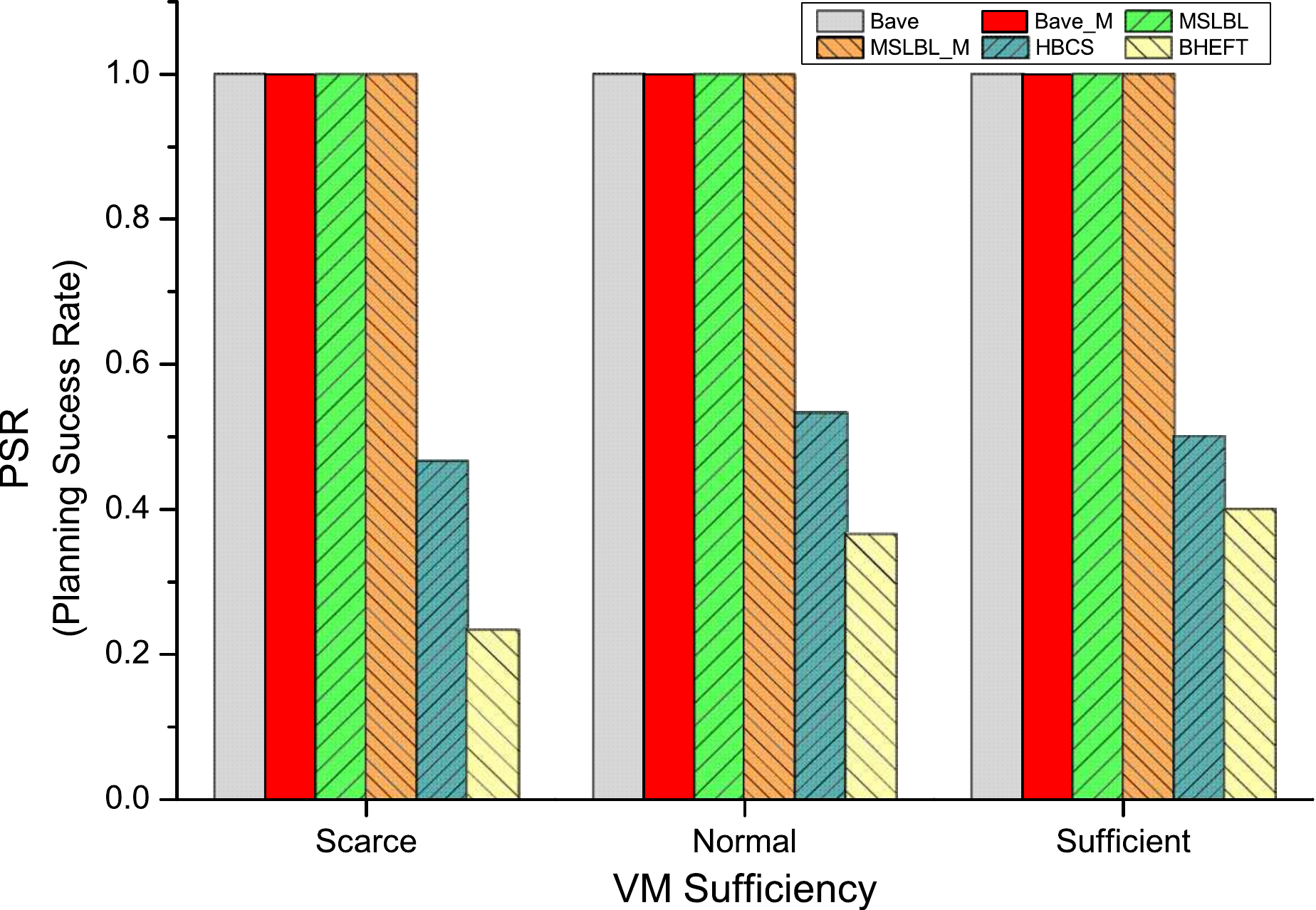} \\ (a)
\end{center}
\end{minipage}
\begin{minipage}{3.5in}
\begin{center}
\includegraphics[width=3.5in]{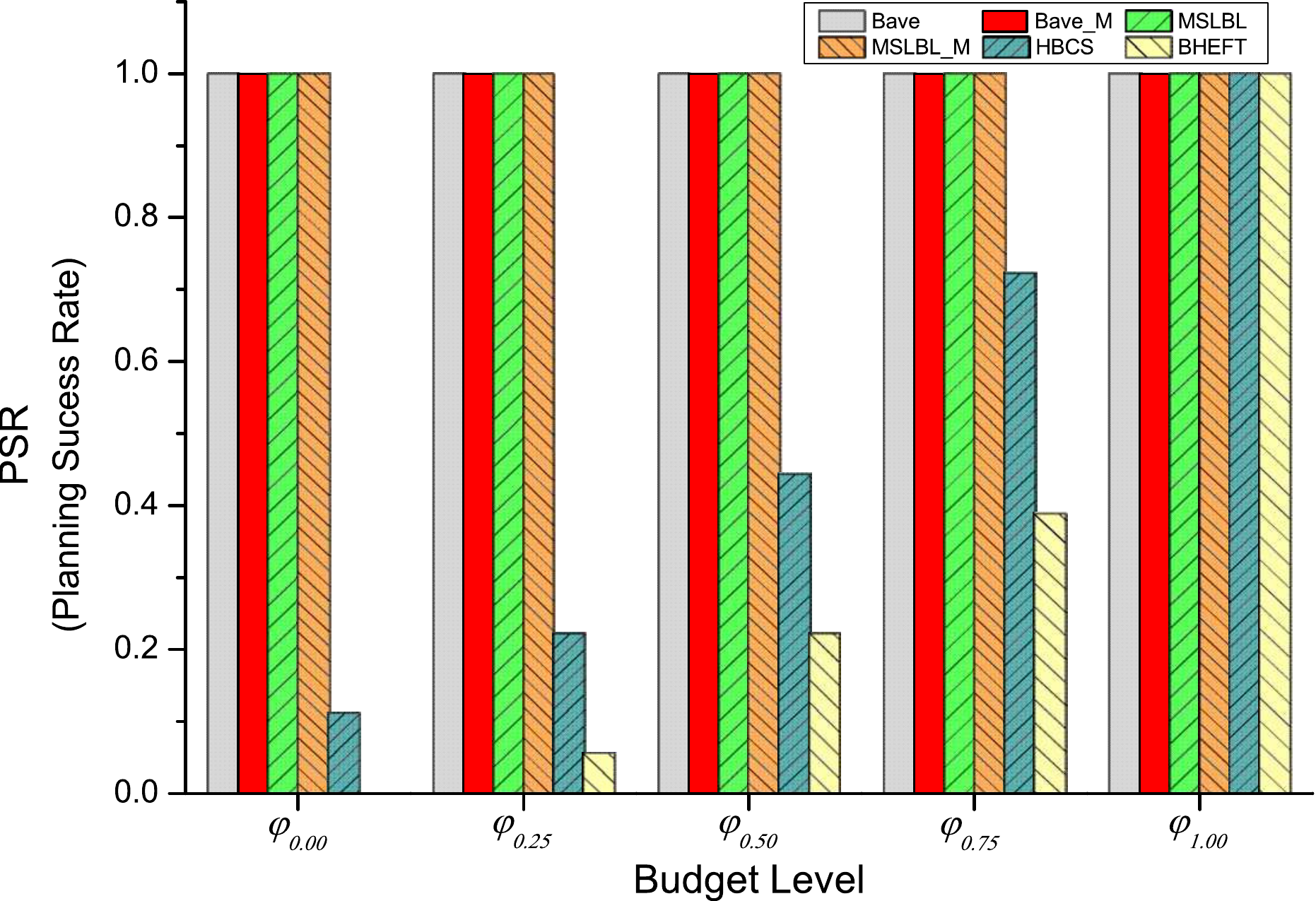} \\ (b)
\end{center}
\end{minipage}
\end{center}
\caption{The scheduling success rates of different algorithms with FFT workflows.}
\label{fig:fft_psr}
\end{figure}

\subsubsection{Success rate of finding a schedule}
Algorithms HBCS and BHEFT cannot find solutions when the budget is limited or
the available VMs are limited for all the workflows we tested.
The other four algorithms can always produce a solution. In Fig. \ref{fig:fft_psr}, we plot the scheduling
success rates for the FFT workflows. The success rates of other workflows have a similar pattern.



\subsection{Multiple workflows}
We conducted tests on multiple workflows. For each workflow type discussed in Section \ref{subset:workflow_setup},
we generate a workflow with at least $1000$ jobs. All workflows are combined to create a mixed set with $N = 9744$ jobs.
In the test, we vary the number of VMs and test four algorithms: BAVE, BAVE\_M, round robin, and random. For the random strategy, we run $1000$ random tests for each test case and report the
average results.
The achieved makespan is summarized in Table \ref{table:multiple_scarce}, \ref{table:multiple_normal}
and \ref{table:multiple_sufficient}. The results show that with more flexible budgets, the BAVE and BAVE\_M algorithms achieve better makespan than the round robin and random strategies. The BAVE\_M algorithm
outperforms the plain BAVE algorithm in most cases.

\begin{table}[H]
\centering
\caption{Multiple large-sized workflows; VM Sufficiency: {\em Scarce}, $N = 9744$}
\begin{tabular}{|c|c|c|c|c|}
\hline
$\varphi$ & BAVE & BAVE\_M & ROUND ROBIN & RANDOM \\
\hline
0.0   &10334   &10334	&10334	 &10334	\\
\hline
0.25   &7083	   &6912   &7114   &7244.06 \\
\hline
0.50  &3740	   &3385   &3876   &4219.71 \\  		
\hline
0.75  &1018	   &1936   &1788   &2124.2 \\  			
\hline
1.00  &878     &855	   &879    &1091.17\\    			
\hline
\end{tabular}
\label{table:multiple_scarce}
\end{table}

\begin{table}[H]
\centering
\caption{Multiple large-sized workflows; VM Sufficiency: {\em Normal}, $N = 9744$}
\begin{tabular}{|c|c|c|c|c|}
\hline
$\varphi$ & BAVE & BAVE\_M & ROUND ROBIN & RANDOM \\
\hline
0.0	  &8791	  &8791	  &8791   &8791	\\
\hline
0.25   &6015	  &5843   &6400   &6344.45 \\
\hline
0.50  &2908	  &3195	  &3342   &3211.83 \\  		
\hline
0.75  &834    &815	  &1278   &1348.84 \\ 		
\hline
1.00  &815	  &815	  &891    &894.06 \\   			
\hline
\end{tabular}
\label{table:multiple_normal}
\end{table}

\begin{table}[H]
\centering
\caption{Multiple large-sized workflows; VM Sufficiency: {\em Sufficient}, $N = 9744$}
\begin{tabular}{|c|c|c|c|c|}
\hline
$\varphi$ & BAVE & BAVE\_M & ROUND ROBIN & RANDOM \\
\hline
0.0	 &6989	&6989	&6989   &6989 \\
\hline
0.25	 &4971	&4616	&4447   &4709.12 \\
\hline
0.50   &1900	&1707	&2195   &2179.32 \\  	 		
\hline
0.75    &725	&695	&970	&984.96 \\  				
\hline
1.00  &725	&695	&777    &777.74 \\ 		
\hline
\end{tabular}
\label{table:multiple_sufficient}
\end{table}

\section{Conclusion}
\label{sec:conclusion}

DAG-based complex workflows are becoming significant workload in the cloud.
In scheduling workflows, the budget constraint is an important factor of consideration due to the pay-as-you-go nature of the cloud.
In this paper, we formulate the workflow scheduling problem with
budget constraints as an integer programming model.
Improving upon the plain upward-rank priority scheme, we propose a weighted scheme using the stationary probabilities of a random walk on the digraph as the weights.
We further design a uniform spare budget splitting strategy, which assigns the spare budget uniformly across all the jobs.
The empirical results show that the uniform spare budget splitting scheme outperforms the earlier scheme that splits the spare budget in
proportion to extra demand, and the weighted priority scheme further improves the workflow makespan.
The advantage of the weighted priority scheme is due to its ability to evaluate the jobs' global importance in the workflow, by considering not only the jobs on the critical path but also off the critical path. Because of the diversity and complexity of workflow types in production, there may be some other unknown factors yet to be studied. Deep analysis of the structural characteristics of different workflows may lead to some new discovery and help design a further improved task priority assignment strategy. For instance, we can borrow the idea proposed in LDCP \cite{Daoud2008} that assigns a higher priority to a job with more children whenever there is a tie. These kinds of refinement that relies on deep analysis of the workflow topologies will be a direction of future research.

%
%
%
%



\section*{Acknowledgment}
This work was supported by the Shanghai Committee of Science and Technology, China (Grant No. 14510722300, 18DZ2203900).

\ifCLASSOPTIONcaptionsoff
  \newpage
\fi



%
%
%

\bibliographystyle{IEEEtran}
\bibliography{./workflow}

\begin{thebibliography}{10}
\providecommand{\url}[1]{#1}
\csname url@samestyle\endcsname
\providecommand{\newblock}{\relax}
\providecommand{\bibinfo}[2]{#2}
\providecommand{\BIBentrySTDinterwordspacing}{\spaceskip=0pt\relax}
\providecommand{\BIBentryALTinterwordstretchfactor}{4}
\providecommand{\BIBentryALTinterwordspacing}{\spaceskip=\fontdimen2\font plus
\BIBentryALTinterwordstretchfactor\fontdimen3\font minus
  \fontdimen4\font\relax}
\providecommand{\BIBforeignlanguage}[2]{{%
\expandafter\ifx\csname l@#1\endcsname\relax
\typeout{** WARNING: IEEEtran.bst: No hyphenation pattern has been}%
\typeout{** loaded for the language `#1'. Using the pattern for}%
\typeout{** the default language instead.}%
\else
\language=\csname l@#1\endcsname
\fi
#2}}
\providecommand{\BIBdecl}{\relax}
\BIBdecl

\bibitem{Juve2012}
G.~Juve, E.~Deelman, G.~B. Berriman, B.~P. Berman, and P.~Maechling, ``An
  evaluation of the cost and performance of scientific workflows on amazon
  {EC2},'' \emph{J. Grid Comput.}, vol.~10, no.~1, pp. 5--21, Mar. 2012.

\bibitem{Wang2014}
Y.~Wang and W.~Shi, ``Budget-driven scheduling algorithms for batches of
  mapreduce jobs in heterogeneous clouds,'' \emph{IEEE Transactions on Cloud
  Computing}, vol.~2, no.~3, pp. 306--319, July 2014.

\bibitem{Rodriguez2014}
M.~A. Rodriguez and R.~Buyya, ``Deadline based resource provisioningand
  scheduling algorithm for scientific workflows on clouds,'' \emph{IEEE
  Transactions on Cloud Computing}, vol.~2, no.~2, pp. 222--235, April 2014.

\bibitem{LENSTRA1977343}
J.~Lenstra, A.~R. Kan, and P.~Brucker, ``Complexity of machine scheduling
  problems,'' in \emph{Studies in Integer Programming}, ser. Annals of Discrete
  Mathematics, P.~Hammer, E.~Johnson, B.~Korte, and G.~Nemhauser, Eds.\hskip
  1em plus 0.5em minus 0.4em\relax Elsevier, 1977, vol.~1, pp. 343 -- 362.

\bibitem{Schulz1996}
A.~S. Schulz, ``Scheduling to minimize total weighted completion time:
  Performance guarantees of lp-based heuristics and lower bounds,'' in
  \emph{Integer Programming and Combinatorial Optimization}, W.~H. Cunningham,
  S.~T. McCormick, and M.~Queyranne, Eds.\hskip 1em plus 0.5em minus
  0.4em\relax Berlin, Heidelberg: Springer Berlin Heidelberg, 1996, pp.
  301--315.

\bibitem{Topcuoglu02}
H.~Topcuoglu, S.~Hariri, and M.-Y. Wu, ``Performance-effective and
  low-complexity task scheduling for heterogeneous computing,'' \emph{IEEE
  Transactions on Parallel and Distributed Systems}, vol.~13, no.~3, pp.
  260--274, Mar 2002.

\bibitem{Gurobi}
\emph{Gurobi Optimization: The state-of-the-art mathematical programming solver
  for prescriptive analytics}, Gurobi, http://www.gurobi.com/, accessed on
  20.09.2018.

\bibitem{Kwok99}
Y.-K. Kwok and I.~Ahmad, ``Static scheduling algorithms for allocating directed
  task graphs to multiprocessors,'' \emph{ACM Comput. Surv.}, vol.~31, no.~4,
  pp. 406--471, Dec. 1999.

\bibitem{Daoud2008}
M.~I. Daoud and N.~Kharma, ``A high performance algorithm for static task
  scheduling in heterogeneous distributed computing systems,'' \emph{Journal of
  Parallel and distributed computing}, vol.~68, no.~4, pp. 399--409, 2008.

\bibitem{Xiang2017}
B.~Xiang, B.~Zhang, and L.~Zhang, ``Greedy-ant: ant colony system-inspired
  workflow scheduling for heterogeneous computing,'' \emph{IEEE Access},
  vol.~5, pp. 11\,404--11\,412, 2017.

\bibitem{Shu2017}
T.~Shu and C.~Q. Wu, ``Performance optimization of {Hadoop} workflows in public
  clouds through adaptive task partitioning,'' in \emph{IEEE INFOCOM 2017 -
  IEEE Conference on Computer Communications}, May 2017, pp. 1--9.

\bibitem{Chen2017}
W.~Chen, G.~Xie, R.~Li, Y.~Bai, C.~Fan, and K.~Li, ``Efficient task scheduling
  for budget constrained parallel applications on heterogeneous cloud computing
  systems,'' \emph{Future Gener. Comput. Syst.}, vol.~74, no.~C, pp. 1--11,
  Sep. 2017.

\bibitem{Sakellariou2007}
R.~Sakellariou, H.~Zhao, E.~Tsiakkouri, and M.~D. Dikaiakos, \emph{Scheduling
  workflows with budget constraints}.\hskip 1em plus 0.5em minus 0.4em\relax
  Boston, MA: Springer US, 2007, pp. 189--202.

\bibitem{Zheng2013}
W.~Zheng and R.~Sakellariou, ``Budget-deadline constrained workflow planning
  for admission control,'' \emph{Journal of Grid Computing}, vol.~11, no.~4,
  pp. 633--651, Dec 2013.

\bibitem{Arabnejad2014}
H.~Arabnejad and J.~G. Barbosa, ``A budget constrained scheduling algorithm for
  workflow applications,'' \emph{Journal of Grid Computing}, vol.~12, no.~4,
  pp. 665--679, Dec 2014.

\bibitem{ABRISHAMI2013}
S.~Abrishami, M.~Naghibzadeh, and D.~H. Epema, ``Deadline-constrained workflow
  scheduling algorithms for infrastructure as a service clouds,'' \emph{Future
  Generation Computer Systems}, vol.~29, no.~1, pp. 158 -- 169, 2013, including
  Special section: AIRCC-NetCoM 2009 and Special section: Clouds and
  Service-Oriented Architectures.

\bibitem{Sahni2018}
J.~Sahni and D.~P. Vidyarthi, ``A cost-effective deadline-constrained dynamic
  scheduling algorithm for scientific workflows in a cloud environment,''
  \emph{IEEE Transactions on Cloud Computing}, vol.~6, no.~1, pp. 2--18, Jan
  2018.

\bibitem{Zhao2006}
H.~Zhao and R.~Sakellariou, ``Scheduling multiple {DAGs} onto heterogeneous
  systems,'' in \emph{Proceedings of the 20th International Conference on
  Parallel and Distributed Processing}, ser. IPDPS'06, 2006, pp. 159--159.

\bibitem{Xie2017}
X.~Guoqi, L.~Liangjiao, Y.~Liu, and L.~Renfa, ``Scheduling trade-off of dynamic
  multiple parallel workflows on heterogeneous distributed computing systems,''
  \emph{Concurrency and Computation: Practice and Experience}, vol.~29, no.~2,
  p. e3782, 2017.

\bibitem{Rodriguez2018}
M.~A. Rodriguez and R.~Buyya, ``Scheduling dynamic workloads in multi-tenant
  scientific workflow as a service platforms,'' \emph{Future Generation
  Computer Systems}, vol.~79, pp. 739 -- 750, 2018.

\bibitem{Wang2017}
Y.~Wang, Y.~Xia, and S.~Chen, ``Using integer programming for workflow
  scheduling in the cloud,'' in \emph{2017 IEEE 10th International Conference
  on Cloud Computing (CLOUD)}, June 2017, pp. 138--146.

\bibitem{Meena2016}
J.~Meena, M.~Kumar, and M.~Vardham, ``Cost effective genetic algorithm for
  workflow scheduling in cloud under deadline constraint,'' \emph{IEEE Access},
  vol.~4, pp. 5065--5082, 2016.

\bibitem{Li2018}
W.~Li, Y.~Xia, M.~Zhou, X.~Sun, and Q.~Zhu, ``Fluctuation-aware and predictive
  workflow scheduling in cost-effective infrastructure-as-a-service clouds,''
  \emph{IEEE Access}, vol.~6, pp. 61\,488--61\,502, 2018.

\bibitem{Zhou2016}
A.~C. Zhou, B.~He, and C.~Liu, ``Monetary cost optimizations for hosting
  workflow-as-a-service in {IaaS} clouds,'' \emph{IEEE Transactions on Cloud
  Computing}, vol.~4, no.~1, pp. 34--48, Jan 2016.

\bibitem{Zheng2016}
Z.~Zheng and N.~B. Shroff, ``Online multi-resource allocation for deadline
  sensitive jobs with partial values in the cloud,'' in \emph{IEEE INFOCOM 2016
  - The 35th Annual IEEE International Conference on Computer Communications},
  April 2016, pp. 1--9.

\bibitem{Champati2017}
J.~P. Champati and B.~Liang, ``Efficient minimization of sum and differential
  costs on machines with job placement constraints,'' in \emph{IEEE INFOCOM
  2017 - IEEE Conference on Computer Communications}, May 2017, pp. 1--9.

\bibitem{Mao2011}
M.~Mao and M.~Humphrey, ``Auto-scaling to minimize cost and meet application
  deadlines in cloud workflows,'' in \emph{International Conference for High
  Performance Computing, Networking, Storage and Analysis}, November 2011, pp.
  1--12.

\bibitem{Silva2014}
R.~F. da~Silva, W.~Chen, G.~Juve, K.~Vahi, and E.~Deelman, ``Community
  resources for enabling research in distributed scientific workflows,'' in
  \emph{2014 IEEE 10th International Conference on e-Science (eScience 2014)},
  2014.

\bibitem{EC2Inst}
Amazon, ``{Amazon EC2 Instances},'' ,
  http://aws.amazon.com/ec2/instance-types/, accessed on 20.09.2018.

\end{thebibliography}

\end{document}